\documentclass[a4paper,11pt]{article}
\usepackage{jcappub} 
\usepackage{lineno}

\usepackage{paralist}
\usepackage{amsmath,amssymb,amsfonts}
\usepackage{verbatim}
\usepackage{bm}
\usepackage{graphicx}
\usepackage{rotating}
\usepackage{float}
\usepackage{xcolor, soul}
\usepackage{hyperref}
\usepackage{subfigure}
\usepackage[normalem]{ulem}
\usepackage[bottom]{footmisc}
\usepackage{algorithm}
\usepackage{algorithmic}

\newcommand{\dd}{\mathrm{d}}

\title{}

\DeclareRobustCommand{\okina}{%
  \raisebox{\dimexpr\fontcharht\font`A-\height}{%
    \scalebox{0.8}{`}%
  }%
}
\pdfoutput=1
\allowdisplaybreaks

\title{Cosmology and Astrophysics of CP-Violating Axions}

\author[a]{Omar F.~Ramadan,}
\author[a]{Jeremy Sakstein,}
\affiliation[a]{Department of Physics \& Astronomy,
University of Hawai\okina i at M\=anoa,
Watanabe Hall, 2505 Correa Road, Honolulu, HI, 96822, USA}
\emailAdd{oramadan@hawaii.edu}
\emailAdd{sakstein@hawaii.edu}

\author[b]{and Djuna Croon} \emailAdd{djuna.l.croon@durham.ac.uk}
\affiliation[b]{Institute for Particle Physics Phenomenology, Department of Physics, Durham University, Durham DH1 3LE, U.K.}

\date{\today}
\abstract{We study the cosmology and astrophysics of axion-like particles (ALPs) with CP-violating Yukawa couplings to nucleons.~At finite nucleon density, the ALP's dynamics is governed by an effective potential which is the sum of the bare periodic potential and a linear potential whose strength depends on the nucleon density.~We identify a \textit{critical nucleon density} $\rho_c$ controlling the dynamics.~At densities smaller than $\rho_c$ the effective potential is a tilted sinusoidal curve and the field is displaced from its zero-density minimum.~At densities larger than $\rho_c$ the minima (and maxima) are absent, and the ALP is destabilized.~Astrophysically, this implies that neutron stars can source a radial ALP field, providing a complementary probe to equivalence principle tests.~Cosmologically, the ALP may have been destabilized in the early Universe and could have made large field excursions.~We discuss model-building applications of our results for such early universe scenarios.}

\begin{document}

\maketitle
\section{Introduction}

Axions and axion-like particles (ALPs) are shift-symmetric pseudo-scalar particles that manifest in various beyond the Standard Model (BSM) scenarios, encompassing the QCD axion, grand unified theories, and string theory \cite{Svrcek:2006yi,Marsh:2015xka,Chadha-Day:2021szb,Alexander:2023wgk,Apers:2024ffe}.~They represent a potential resolution to the strong CP problem \cite{Peccei:1977hh,Peccei:1977ur,Peccei:2006as,Kim:2008hd} and serve as candidates for dark matter (DM) \cite{Marsh:2015xka,Chadha-Day:2021szb,OHare:2024nmr}.~Their mass and couplings are protected against radiative corrections by a shift symmetry, making them valuable for building natural models of inflation \cite{Freese:1990rb,Pajer:2013fsa,Croon:2014dma,Croon:2015fza}, dark energy (DE) \cite{Frieman:1995pm}, and early dark energy \cite{Poulin:2018cxd,McDonough:2022pku}.~These considerations have motivated a concerted global search effort looking for their signatures in astrophysical objects and terrestrial laboratories (see e.g.,~\cite{Baryakhtar:2022hbu, Adams:2022pbo,DiLuzio:2020wdo,Sikivie:2020zpn} for reviews).~

CP-violating ALP-nucleon Yukawa couplings typically emerge as a result of CP-violation in the UV \cite{Moody:1984ba,Georgi:1986kr,Bertolini:2020hjc,OHare:2020wah,DiLuzio:2021jfy}.~These induce monopole-monopole forces between objects \cite{Moody:1984ba}, providing additional detection channels e.g., via fifth-force and weak equivalence principle violation searches \cite{Barbieri:1996vt,Pospelov:1997uv,OHare:2020wah}.~In this work, we explore the consequences of these couplings for the cosmology of ALPs, and for the existence and structure of neutron stars.~We will consider a generic ALP in order to be model-independent.

\begin{figure}[ht]
    \centering
   \includegraphics[width=0.44\textwidth]{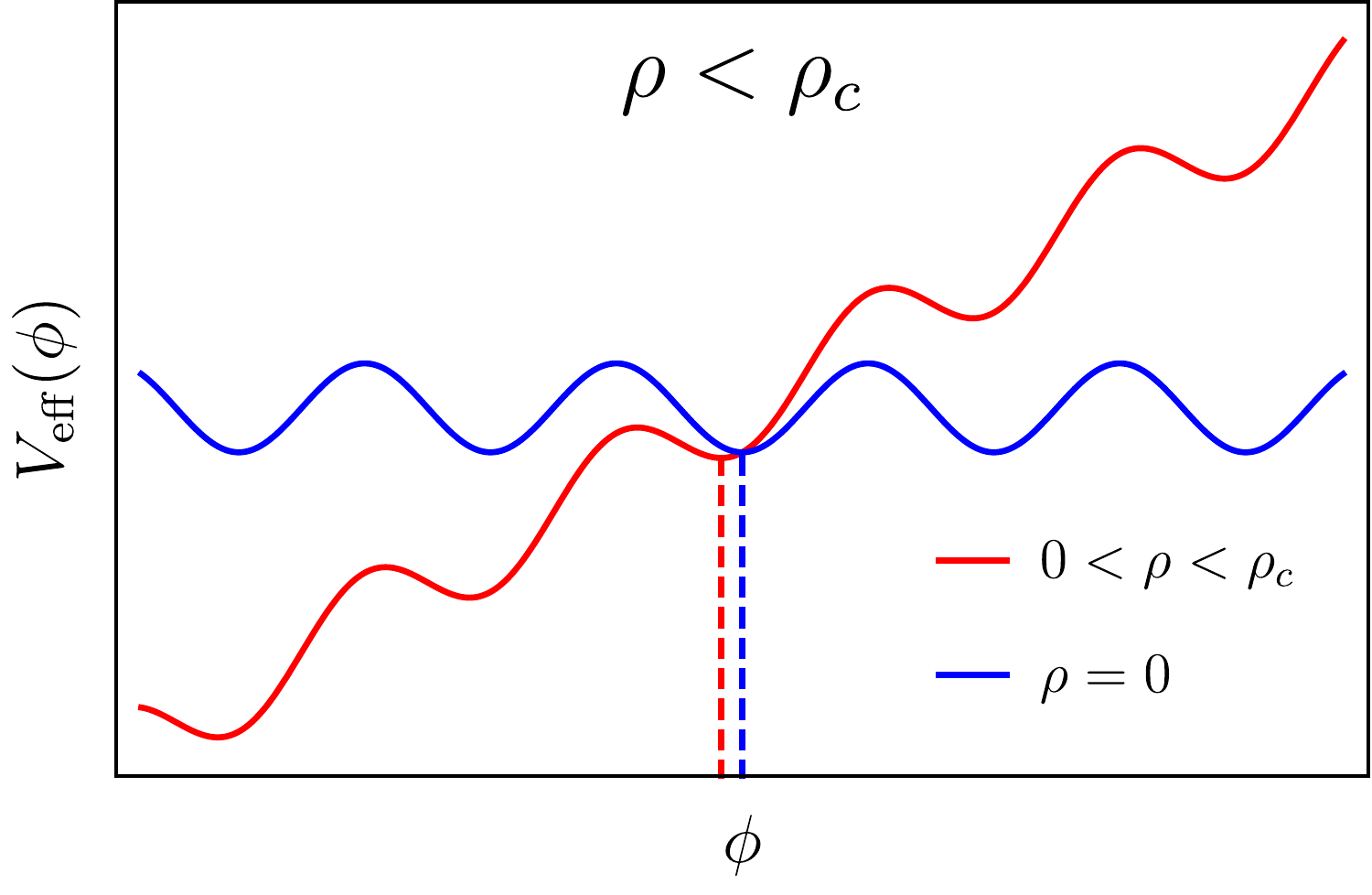}    \includegraphics[width=0.44\textwidth]{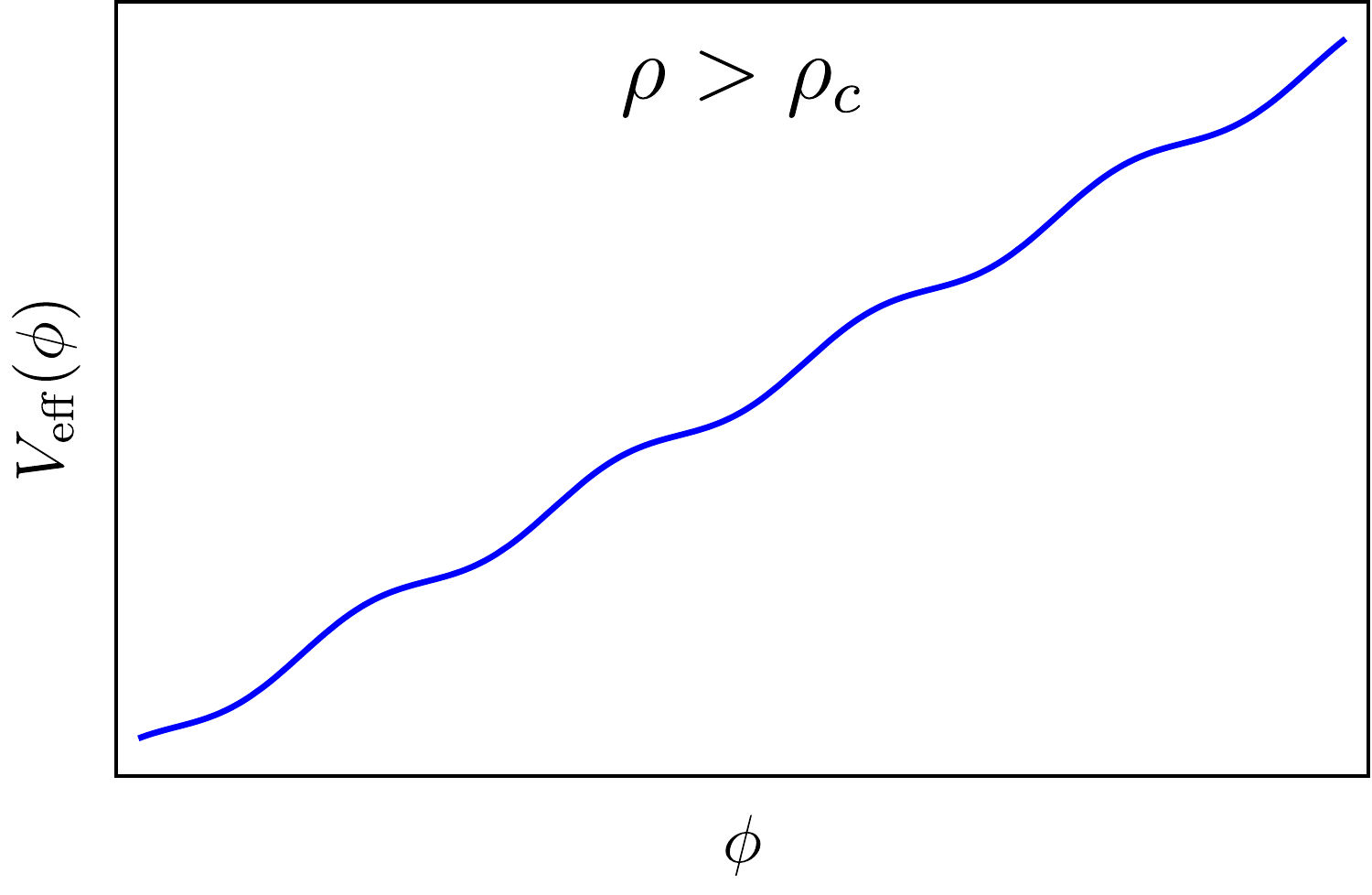}
    \caption{\textbf{Left:}~Effective potential when $\rho<\rho_c$.~The blue curve shows the bare potential ($V_{\rm eff}$ at $\rho=0$) and the red curve shows the effective potential when $0<\rho<\rho_c$.~The dashed lines indicate the minima, demonstrating that the VEV is shifted to smaller values at finite density.~\textbf{Right:}~Effective potential when $\rho>\rho_c$.~In this case, there are no minima.}
    \label{fig:tilting}
\end{figure}

Instanton corrections  break the ALP shift-symmetry $a(x)\rightarrow a(x)+a_0$ with $a_0$ an arbitrary constant to a discrete symmetry $a(x)\rightarrow a(x)+2\pi n$ with $n$ an integer.~This generically generates a periodic potential of the form $V(a)=m_a^2f_a^2[1-\cos(a/f_a)]$ for a single ALP field \cite{Coleman:1985rnk,Svrcek:2006yi,Csaki:1998vv} where $m_a$ is the ALP's mass and $f_a$ is the decay constant.~As we will derive below, the ALP-nucleon coupling induces a density-dependent effective potential $V_{\rm eff}(a)=V(a)+\rho a/\mu$ where $\rho$ is the density of nucleons and $\mu$ is a new mass-scale parameterizing the CP-violating coupling.~The effect of the linear term is to tilt the potential in an environment-dependent manner that depends upon the \textit{critical density} $\rho_c=m_a^2\mu f_a$;~this is exemplified in Fig.~\ref{fig:tilting}.~When $\rho<\rho_c$ the minima are displaced, shifting the ALP vacuum expectation value (VEV) to smaller values where the effective mass is lighter i.e., $m_{\rm eff}=V''_{\rm eff}(a)<m_a$.~When $\rho>\rho_{c}$ the minima are absent, and the ALP is destabilized.

Cosmologically, the density of nucleons redshifts so the ALP may be destabilized at early times.~If this is the case, then at some later time the nucleon density will redshift sufficiently for the minima to appear, with the specific transition time being parameter-dependent.~After this, the field will evolve towards the nearest minimum.~Astrophysically, one expects that the ALP VEV is shifted inside celestial objects.~This implies that its mass is lighter than it is in interstellar space, increasing the range of the ALP-mediated fifth-force.~Depending on the size of $\mu$, sufficiently dense objects may destabilize the ALP.~In what follows, we explore these possibilities quantitatively.~To summarize our results, we find:
\begin{itemize}
    \item[\textbf{Cosmological ALPs:}] \mbox{}
    \begin{itemize}
        \item In the stabilized regime the ALP exhibits oscillations about a time-dependent VEV.~The former acts as a component of DM while the latter acts as DE.~Thus, CP-violating ALPs provide a unified description of these phenomena.
        \item At early times, the ALP is destabilized and we derive a bound on the model parameters that must be satisfied in order that the nucleon masses do not vary significantly between big bang nucleosynthesis (BBN) and the cosmic microwave background (CMB).
    \end{itemize}
    
    \item[\textbf{Astrophysical ALPs:}]\mbox{}
    \begin{itemize}
        \item The ALP may be destabilized inside neutron stars, causing it to make large field excursions that would unbind the star.~We derive bounds on the model parameters that ensure that this cannot happen so that these objects exist.
        \item The ALP can be probed by solar system tests of gravity.~We derive the corresponding bounds on the model parameters.
    \end{itemize}
\end{itemize}

This work is organized as follows.~We introduce CP-violating ALPs and derive the features outlined above in section~\ref{sec:formalism}.~The consequences of the CP-violating couplings for the ALP's cosmological evolution are derived in section~\ref{sec:cosmology} and the consequences for neutron stars in section~\ref{sec:NS}.~We discuss the implications of our results and conclude in section~\ref{sec:discussion}.

\section{CP-Violating Axions}
\label{sec:formalism}

The low-energy effective ALP action, valid at sub-MeV energy scales is 
\begin{align}
\label{eq:ALP_Lagrangian}
S_a&=\int\dd^4x\sqrt{-g}\left\{-\frac12\partial_\mu a\partial^\mu a- m_a^2f_a^2\left[1-\cos\left(\frac{a}{f_a}\right)\right]\right.\nonumber\\&\left.+\sum_{\psi=n,p}\left(i\bar{\psi}\gamma^\mu\partial_\mu\psi-m_\psi\left(1+\frac{a}{\mu_\psi}\right)\bar{\psi}\psi\right)+\cdots\right\},
\end{align}
where $\psi$ are Dirac spinors corresponding to either neutrons ($\psi=n$) or protons ($\psi=p$), $\mu_\psi$ are new mass scales parameterizing the strength of the CP-violating Yukawa ALP-nucleon  interactions, and $\cdots$ represent other ALP interactions and higher order terms not relevant for this study.~The Yukawa terms are sometimes written equivalently as $-a\sum_\psi g^\psi_s\bar{\psi}\psi$ with $\psi=n,p$~e.g., \cite{OHare:2020wah};~the relation is $g^\psi_s=m_\psi/\mu_\psi$ as we derive in Appendix~\ref{app:MG}.~In what follows we will take $\mu_n=\mu_p=\mu$ or, equivalently, $g_s^n=g_s^p=g_s^N$ ($N$ is for nucleon) for simplicity, and to enable comparisons with previous works who have made the same choice e.g., reference \cite{OHare:2020wah}.~The action we consider is then
\begin{align}
    \label{eq:ALP_Lagrangian_simple}
S_a&=\int\dd^4x\sqrt{-g}\left\{-\frac12\partial_\mu a\partial^\mu a- m_a^2f_a^2\left[1-\cos\left(\frac{a}{f_a}\right)\right]\right.\nonumber\\&\left.+i\bar{n}\gamma^\mu\partial_\mu n+i\bar{p}\gamma^\mu\partial_\mu p-m_n\left(1+\frac{a}{\mu}\right)\bar{n}n-m_p\left(1+\frac{a}{\mu}\right)\bar{p}p\right\} .
\end{align}
This action is equivalent to a conformal scalar-tensor theory of gravity \cite{Wetterich:2014bma,Burrage:2018dvt} where the \textit{Einstein frame metric} $g_{\mu\nu}$ satisfies the Einstein equations and neutrons and protons move on geodesics of the \textit{Jordan frame metric} $\tilde{g}_{\mu\nu}=\exp(a/\mu)g_{\mu\nu}$;~see e.g., \cite{Sakstein:2013pda,Sakstein:2014jrq,Burrage:2016bwy,Burrage:2017qrf,Sakstein:2017pqi,Sakstein:2018fwz,Baker:2019gxo,Brax:2021wcv} for reviews.~We provide a brief review of this equivalence in Appendix~\ref{app:MG}.~As a consequence of this coupling, the field's equation of motion (EOM) is 
\begin{equation}
    \label{eq:fiedlEOM_General}
    \Box a-\frac{\dd V_{\rm eff}(a)}{\dd a}=0;\quad\,V_{\rm eff}(a)=V(a)-\frac{a}{\mu}T,
\end{equation}
where $V_{\rm eff}$ is the \textit{effective potential} that governs the field's dynamics, $\Box=g^{\mu\nu}\nabla_\mu\nabla_\nu$, and $T=g_{\mu\nu}T^{\mu\nu}$ is the trace of the energy-momentum tensor for protons and neutrons.~In cases where the pressure can be neglected, including non-relativistic sources and cosmology, the effective potential is
\begin{equation}
    V_{\rm eff}(a)=m_a^2f_a^2\left[1-\cos\left(\frac{a}{f_a}\right)\right]+\frac{\rho}{\mu}a,
    \label{eq:Veff}
\end{equation}
where $\rho=\rho_n+\rho_p$ is the density of protons and neutrons.~The effective potential has an infinite number of minima labeled by an integer $n$
\begin{equation}
    \label{eq:phiMinALP}
    a_{\rm min}=-f_a \arcsin{\left[\frac{\rho}{\rho_{c}}\right]}+ 2\pi n f_a,
\end{equation}
when $\rho<\rho_c$, where $\rho_c\equiv\mu f_a m_a^2$ is the \textit{critical density}.~In contrast, $V_{\rm eff}(a)$ has no minima (or maxima) when $\rho>\rho_c$.~The two scenarios are exemplified in Fig.~\ref{fig:tilting}.~As demonstrated in this figure by the vertical dashed lines, the position of the minima is shifted for nonzero but sub-critical density, and the field's mass at each minimum is
\begin{equation}
    \label{eq:meff}
    m^2_{\rm eff}=V_{\rm eff}''(a_{\rm min})=m_a^2\sqrt{1-\left(\frac{\rho}{\rho_{c}}\right)^2}.
\end{equation}
This implies that the CP-violating couplings act to reduce the ALP mass in high-density environments so that the monopole-monopole force range is increased.~For relativistic objects, the picture above is qualitatively unchanged but $\rho$ is replaced by $-T=\rho-3P$ with $P$ the pressure of protons and neutrons.~The majority of physical nuclear equations of state have $\rho-3P>0$, but for the few that do not, the potential in Fig.~\ref{fig:tilting} is tilted the opposite direction.~In this case, the qualitative considerations above still apply, but the field rolls to larger values when $\rho>\rho_c$.

The ALP EFT in equation~\eqref{eq:ALP_Lagrangian_simple} ceases to be a valid description at energies/temperatures larger than $\sim$MeV.~In this case, one must use the UV completion where the ALP-matter couplings are treated at the level of individual quarks.~These too have CP-violating couplings \cite{Moody:1984ba}.~We will remain agnostic of these in what follows in order to be as model-independent as possible, but some general considerations can be discerned.~The discussion above is unchanged except for the replacement of $T$ by the trace of the energy-momentum tensor of the quarks, possibly with different couplings $\mu_q$, which would change $aT/\mu\rightarrow a\sum_qT_q/\mu_q$ with the sum running over all quarks.~The ALP-quark EFT does not break down at temperatures larger than the quark masses, however this does not imply that at such temperatures $\rho-3P=0$ such that the contributions to the ALP potential vanish.~This would be true for exactly massless quarks, but the leading-order correction for a massive particle scales as $m_q^2T_q^2$ where $m_q$ and $T_q$ are the quark mass and temperature respectively \cite{Sakstein:2019fmf,CarrilloGonzalez:2020oac,CarrilloGonzalez:2023lma}.~This increases at higher temperatures, acting to destabilize the ALP potential.~

\section{Cosmology}
\label{sec:cosmology}

\begin{figure}
    \centering   \includegraphics[width=\textwidth]{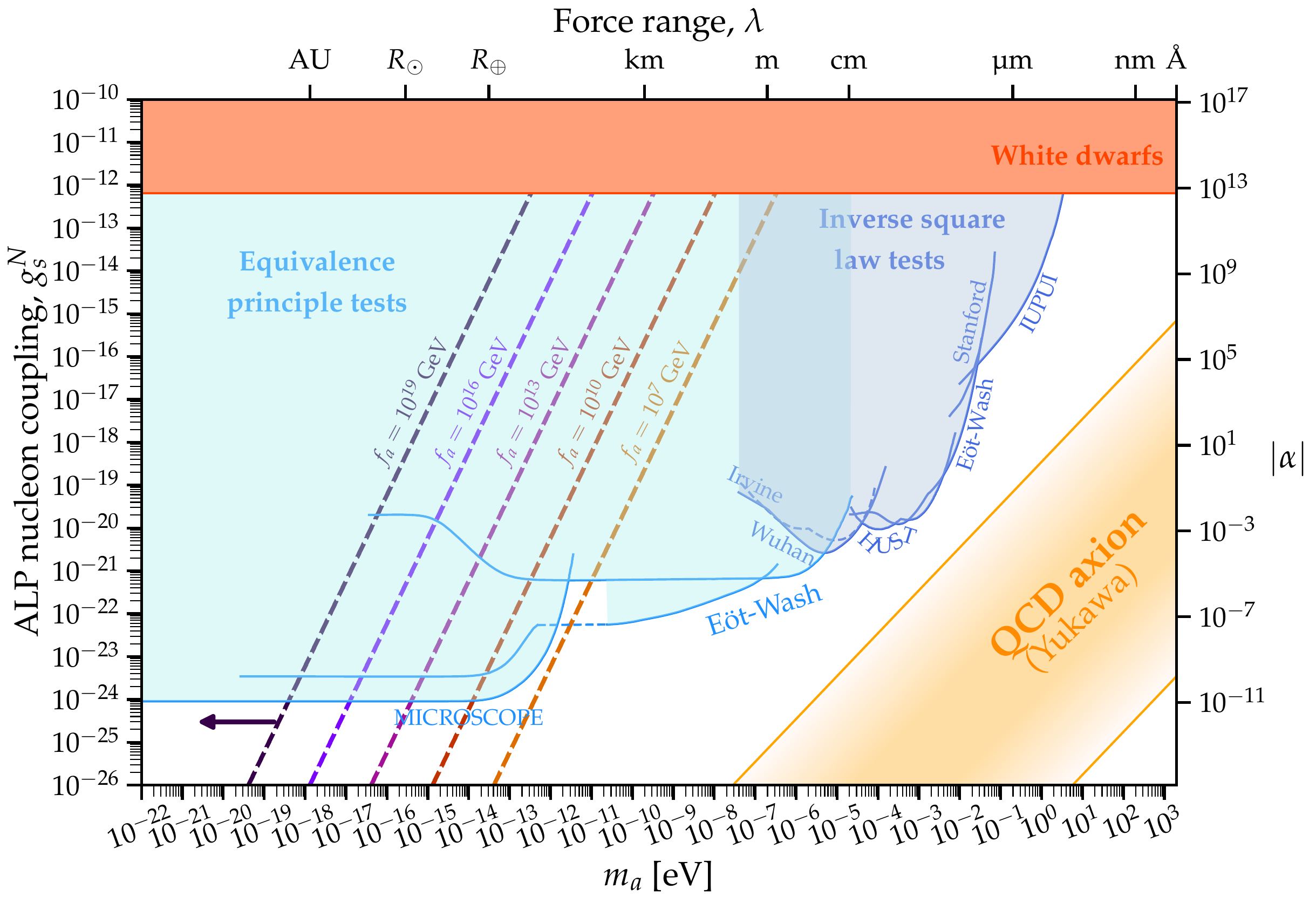}
    \caption{CP-violating ALP parameter space $g_s^N$ vs $m_a$;~we remind the reader that $g_s^N=m_N/\mu$.~The ALP-nucleon Yukawa couplings mediate fifth-forces on length scales given in the top legend with strength relative to gravity $\alpha=F_{\rm ALP}/F_{\rm grav}$ given in the right legend.~The dashed lines show regions (indicated by the arrow) where the ALP is destabilized before the QCD phase transition i.e., $z_c>z_{\rm QCD}$ in Eq.~\eqref{eq:criticalRedshift} for different values of $f_a$ indicated in the figure.~The blue region shows constraints from inverse-square law tests of gravity~\cite{Chen:2014oda,Lee:2020zjt,Kapner:2006si,Tan:2020vpf,Hoskins:1985tn}, the light blue region shows bounds from tests of weak equivalence principle~\cite{Smith:1999cr, Berge:2017ovy}, and the dark orange region shows the astrophysical limits from white dwarf cooling~\cite{Bottaro:2023gep}.~The QCD axion region highlights the expected parameter space for the QCD axion.~The figure was created by modifying the open-source code from reference~\cite{OHare:2020wah}.~
    }
\label{fig:ALP_cosmo_bounds}
\end{figure}

We first study the cosmological implications of ALP-nucleon Yukawa couplings.~In a flat Friedmann-Lemaitre-Robertson-Walker (FLRW) universe, equation~\eqref{eq:fiedlEOM_General} becomes
\begin{equation}
    \label{eq:EOM_Cosmo}
\ddot{a}+3H\dot{a}+m_a^2f_a\sin\left(\frac{a}{f_a}\right)+\frac{\rho_b(t)}{\mu}=0.
\end{equation}
The baryon density $\rho_b(t)=3H^2M_{\rm Pl}^2\Omega_{b,0}(1+z)^3$ increases in the past and therefore, depending on the parameters, we expect that the minima exist at late-times but that the ALP may be destabilized (i.e., $\rho > \rho_c$) in the past at a \textit{critical redshift}
\begin{equation}
\label{eq:criticalRedshift}
   z_c=\left(\frac{\mu f_a m_a^2}{3H_0^2M_{\rm Pl}^2\Omega_{b,0}}\right)^{\frac{1}{3}}-1.
\end{equation}
The effective action \eqref{eq:ALP_Lagrangian_simple} is valid when quarks are confined, below the QCD phase transition at $z_{\rm QCD}\sim10^{12}$, so this expression is accurate provided that $z_c\le z_{\rm QCD}$.~The UV-completion describing the ALP-quark CP-violating couplings is required to treat redshifts larger than this.~As discussed above, the qualitative features are unchanged at high redshift --- the ALP potential will be destabilized at some point in the past.~In our current analysis, we will treat the effects of this as an initial field position and velocity and begin our investigations at $z_{\rm QCD}$ or later.~The regions of parameter space where $z_c>z_{\rm QCD}$ for different values of the axion decay constant are shown in figure~\ref{fig:ALP_cosmo_bounds}.~A general cosmological scenario is as follows.

We expect that the ALP effective potential is destabilized at some point in the past, implying that the cosmology of the ALP begins with a rolling phase after inflation.~This is problematic for some parameters because the field can make large  excursions during this time.~Equation~\eqref{eq:ALP_Lagrangian_simple} predicts that the nucleon masses $m_N$ depend on the ALP VEV as 
\begin{equation}
    \label{eq:nucleon_mass}
    m_N=\bar{m}_N\left(1+\frac{a}{\mu}\right)
\end{equation}
with $\bar{m}_N$ the nucleon mass at $a=0$, which implies a variation in the masses of protons and neutrons whenever the field varies.~Between BBN and the CMB scales, such variations are constrained to be smaller than $\sim10\%$ \cite{Uzan:2010pm}.~In addition, since the kinetic energy of the ALP is expected to increase with time, it may dominate the energy budget of the universe, leading to an early ALP-dominated kination phase.~Finally, the ALP's equation of state $w\ne0$ during this time so if the ALP constitutes DM today then this DM will have been absent in the past, altering the standard $\Lambda$CDM evolution.

To investigate this quantitatively, we will consider the following scenario.~As noted above, prior to the QCD phase transition at $z_{\rm QCD}\sim 10^{12}$ the field is destabilized by ALP-quark couplings so is rolling with some initial position $a_0$ with velocity $\dot{a}_0$.~During the radiation epoch, the ALP's equation of motion  in Eq.~\eqref{eq:EOM_Cosmo} is 
\begin{equation}
    \label{eq:destabilzedRad}
    \ddot{a}+3H\dot{a}=-\frac{\rho_{QCD}}{\mu}\left(\frac{t_{\rm QCD}}{t}\right)^{\frac32},
\end{equation}
where $\rho_{QCD}\approx1.8\times10^{-12} \textrm{ GeV}^4$ is the baryonic density\footnote{We remind the reader that the ALP couples only to nucleons so this value is  smaller than the total density of the universe at the QCD phase transition, $\rho=3H^2M_{\rm pl}^2\sim 5\times10^{-3} \textrm{ GeV}^4$, by several orders of magnitude.~All quoted parameters in this section were calculated using the Planck 2018 results \cite{Planck:2018vyg}.} at the time  of the QCD phase transition, $t_{\rm QCD}\approx2\times10^{-5}$s, and we have neglected the bare potential.~This approximation is valid at early times when the density-dependent term is large, but we expect it to break down after some amount of time has elapsed because the term on the right hand side of Eq.~\eqref{eq:destabilzedRad} redshifts as $t^{-3/2}$.~We can estimate this timescale by comparing the restoring term with the driving term to find 
\begin{equation}
\label{eq:tbreak}
    t_{\rm break}= {\frac{1}{\sqrt{2} m_a}},
\end{equation}
where we have used the fact that $a/f_a\ll1$.~The solution to Eq.~\eqref{eq:destabilzedRad} with the initial conditions $a(t_{\rm QCD})=a_0$ and $\dot{a}_{\rm QCD}=\dot{a}_0$ is
\begin{equation}
    \label{eq:ALP_SOL_RADF_DESTSBILIZED}
    a(t) = a_0+2\dot{a}_0t_{\rm QCD}\left(1-\sqrt{\frac{t_{\rm QCD}}{t}}\right)+\frac{2 \rho_{\rm QCD} t_{\rm QCD}^{3/2}}{\mu }\left(2\sqrt{\frac{t_{\rm QCD}}{t}}-\frac{t_{\rm QCD}}{{t}}-1\right)\sqrt{t},
\end{equation}
demonstrating that at sufficiently late times, the initial conditions $a_0$ and $\dot{a}_0$ become irrelevant and $a(t)\propto-\sqrt{t}$.~Note that the approximations above do not break when the effective potential is stabilized, only when the bare potential becomes important so this solution is valid until the field approaches the nearest minimum.~Taking the asymptotic solution, one finds that the nucleon mass varies by $10\%$ after a time
\begin{equation}
    \label{eq:t_neutron_cosmo}
    t_{N}=\frac{\mu ^4 }{400 \rho_{\rm QCD}^2 t_{\rm QCD}^3}.
\end{equation}
From this, one concludes that $t_{N}\le t_{\rm CMB}$ where $t_{\rm CMB}\approx 380$ kyr is the time of decoupling, when $\mu<5\times10^{18}$ GeV.~This suggests that a constraint on the CP-violating coupling $\mu$ can be placed, but there are caveats that must be addressed before drawing such a conclusion.~First, we have used a series of approximations, one or more of which may break down;~and  second, the bound  only applies to values of $f_a$ and $m_a$ for which the baryon density at the time of decoupling $\rho_{\rm CMB}\approx 2.3\times10^{-36} \textrm{ GeV}^4$ is larger than the critical density $\rho_c$.~When this is not the case the minima reappear before the CMB is formed and the field excursion is smaller than we have estimated.~One can still place a bound in this case provided this effect is accounted for.

\begin{figure}[t]
    \centering
\includegraphics[width=.75\textwidth]{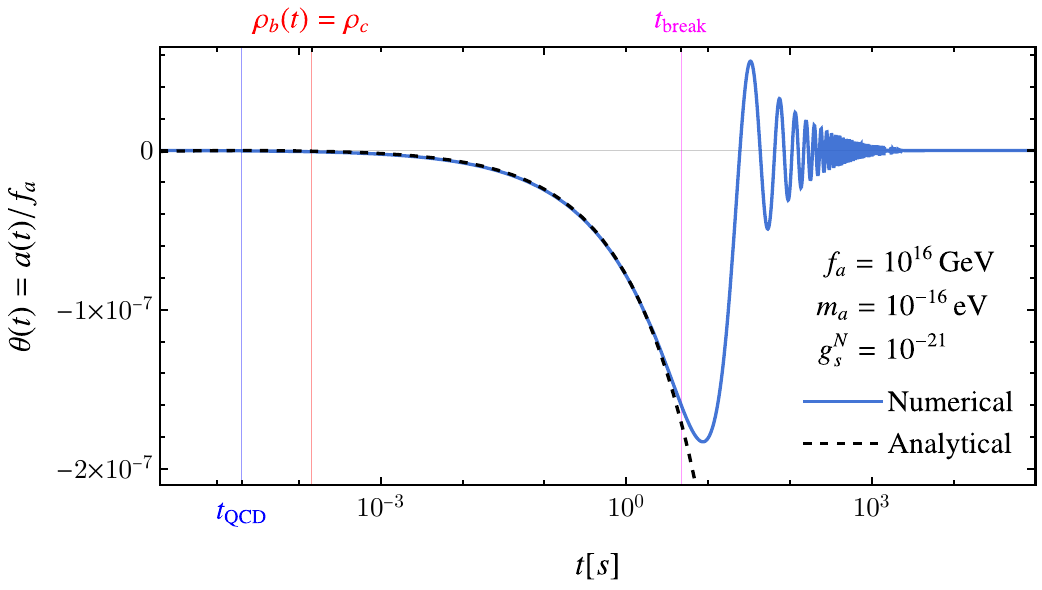} 
    \caption{Cosmological evolution of the ALP field.~The blue solid line is the numerical solution obtained by solving the Klein-Gordon equation in Eq.~\eqref{eq:EOM_Cosmo} and the black dashed line is the approximate analytical solution in Eq.~\eqref{eq:ALP_SOL_RADF_DESTSBILIZED}.~The parameters used were $g_s^N = 10^{-21}$ ($\mu=9.38\times10^{20}$ GeV), $f_a = 10^{16}$ GeV, and $m_a = 10^{-16}$ eV.~The vertical solid red line indicates the time at which the effective potential is stabilized i.e., $\rho(t)=\rho_c$, the vertical solid blue line shows the time of the QCD phase transition, and the vertical solid magenta line shows the estimate for the time $t_{\rm break}$ (Eq.~\eqref{eq:tbreak}) at which the analytic solution in Eq.~\eqref{eq:ALP_SOL_RADF_DESTSBILIZED} is estimated to break down.} 
    \label{fig:ALP_cosmo_destabilized}
\end{figure}

We have further explored the scenario above to investigate these caveats by numerically by solving Eq.~\eqref{eq:EOM_Cosmo} in a radiation dominated universe.~An example is shown in figure~\ref{fig:ALP_cosmo_destabilized}.~The solution rapidly approaches the $-\sqrt{t}$ scaling derived analytically above.~Other choices of parameters gave similar behavior.~We therefore conclude that the analytically derived bound $\mu<5\times10^{18}$ for parameters $m_a$ and $f_a$ where $\rho<\rho_c$ is robust.~This bound is reinforced by tests of the weak equivalence principle (WEP).

As to the question of whether the destabilized ALP can over-close the universe at late times:~equation \eqref{eq:ALP_SOL_RADF_DESTSBILIZED} shows that this is not possible.~The asymptotic solution $a\sim-\sqrt{t}$ implies that the kinetic energy $\dot{a}^2/2$ is decaying while the potential is bounded by virtue of being a cosine.~Ultimately, this can be traced back to the density term in the ALP's EOM decaying as the universe expands.~Interestingly, this implies that there may be a phase of kination in the early universe, although the details of this will depend on the axion-quark coupling, so lie beyond the scope of the present work.

The minima of the effective potential appear at $z=z_c$.~As can be seen in Fig.~\ref{fig:ALP_cosmo_destabilized}, the field will subsequently roll towards the nearest time-dependent minimum, pushed by the density term in Eq.~\eqref{eq:EOM_Cosmo} (which acts as a driving force) and execute harmonic oscillations.~Writing $a(t)=a_{\rm min}(t)+\delta a(t)$ one finds, in the limit where $\rho(t)\ll\rho_c$, appropriate for all times except the short period when $z\approx z_c$:
\begin{equation}
    \ddot{\delta a}+3H\dot{\delta a} +m_{\rm eff}^2 (t)\delta a=0,
\end{equation}
where we have taken $\ddot{a}_{\rm min}\approx\dot{a}_{\rm min}\approx0$.~Thus, the field indeed executes damped harmonic oscillations about $a_{\rm min}(t)$ with period $m_{\rm eff}^{-1}(t)$.~For all intents and purposes one has $m_{\rm eff}\sim m_a$ when $\rho\ll\rho_c$ (see Eq.~\eqref{eq:meff}).~As with regular ALPs, the component $\delta a(t)$ behaves as cold dark matter (CDM) with equation of state $w=0$ \cite{Marsh:2015xka,Chadha-Day:2021szb}.~The component $a_{\rm min}(t)$ induced by the CP-violating couplings behaves as a form of dynamical dark energy with equation of state
\begin{equation}
\label{eq:wEOS}
    w_{\rm DE} = \frac{P_a}{\rho_a}=\frac{\dot{a}_{\rm min}^2-2V(a_{\rm min})}{\dot{a}_{\rm min}^2+2V(a_{\rm min})}=-1+18\frac{H_0^2}{m_a^2}
\end{equation}
where $H_0\sim10^{-33}$ eV is the Hubble constant today, we have again taken the limit $\rho\ll\rho_c$, and the final expression is the value at the present time.~This gives detectable deviations from the $\Lambda$CDM value $w=-1$ for ALPs with masses in the range $m_a\sim10^{-31}$--$10^{-32}$ eV but is negligible for heavier masses.

The case where $m_a\sim H_0\sim 10^{-33}$ eV must be treated separately because the oscillation period is longer than the age of the universe, implying that the field is slowly-rolling towards the nearest minimum and does not act as CDM.~In the slow-roll limit, $\ddot{a}\approx 0$, and taking $\sin(a/f_a)\approx a/f_a$, equation \eqref{eq:EOM_Cosmo} admits the solution
\begin{equation}
    a(t)=a_ie^{-\frac{m_a^2}{4}(t^2-t_i^2)}+\frac{t_0^2\rho_0}{4\mu}e^{-\frac{m_a^2t^2}{4}}\left[\mathrm{Ei}\left(\frac{m_a^2 t_i^2}{4}\right)-\mathrm{Ei}\left(\frac{m_a^2 t^2}{4}\right)\right],
    \label{eq:SR_Solution}
\end{equation}
where $t_i$ is the time at which the field begins to roll;~$a_i=a(t_i)$ is the ALP's initial condition;~and we have taken $H(t)=2/3t$ and $\rho_b=\rho_0(t/t_0)^{2/3}$, corresponding to a matter dominated universe.~The function
\begin{equation}
    \label{eq:def_Ei}
    \mathrm{Ei}(y)=\int_{-\infty}^y\frac{e^{u}}{u}\dd u
\end{equation}
is the exponential integral function.~At late times when $t\gg m_a^{-1}$ we can Taylor-expand Eq.~\eqref{eq:SR_Solution} to find that the equation of state of the field in Eq.~\eqref{eq:wEOS} is 
\begin{equation}
    w_{\rm DE}=-1+\frac{8}{m_a^2t^2}
\end{equation}
so the field acts as a source of dynamical dark energy.

\section{Astrophysical Objects}
\label{sec:NS}

\begin{figure}[t]
    \centering
    \includegraphics[width=0.45\textwidth]{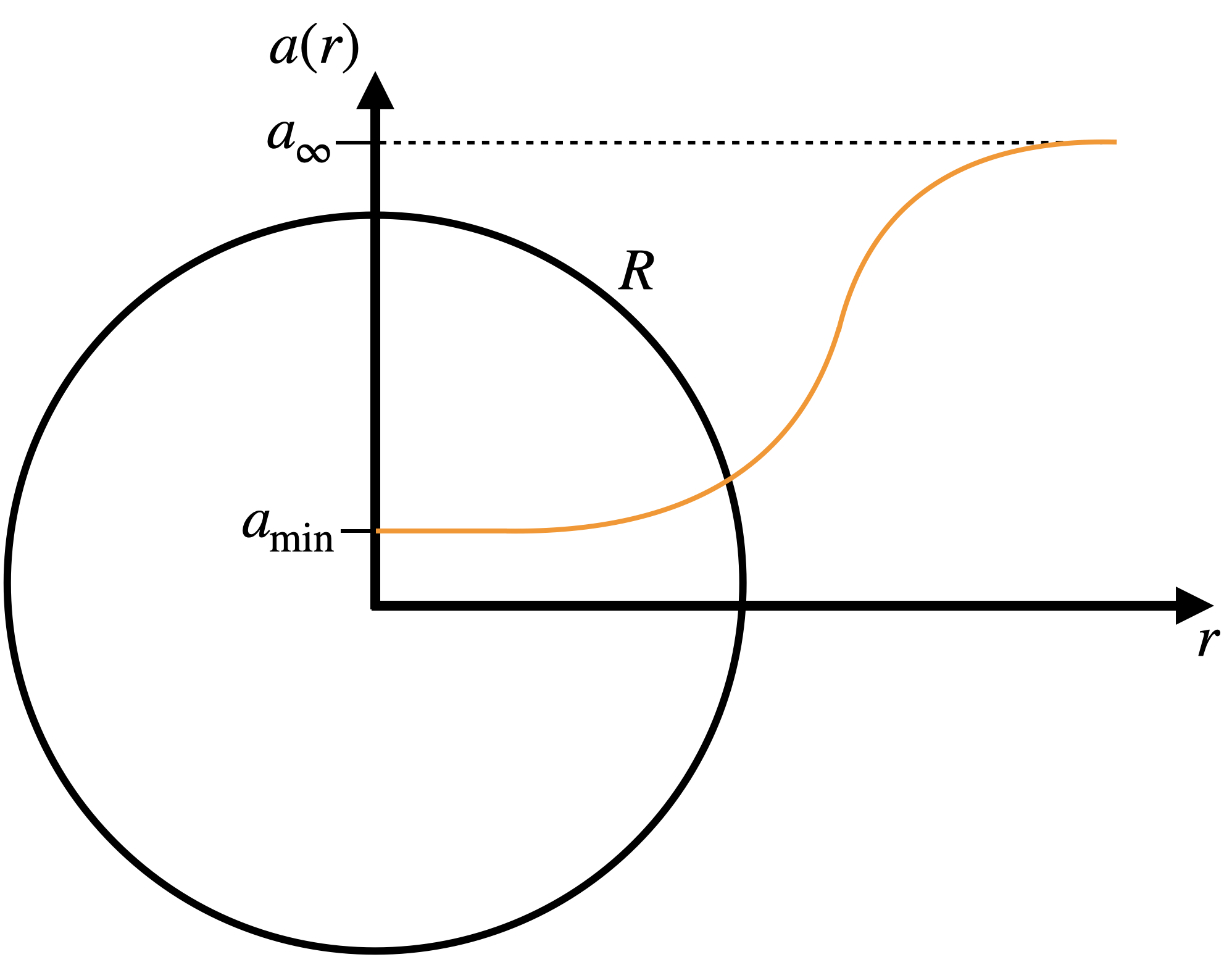}
    \includegraphics[width=0.45\textwidth]{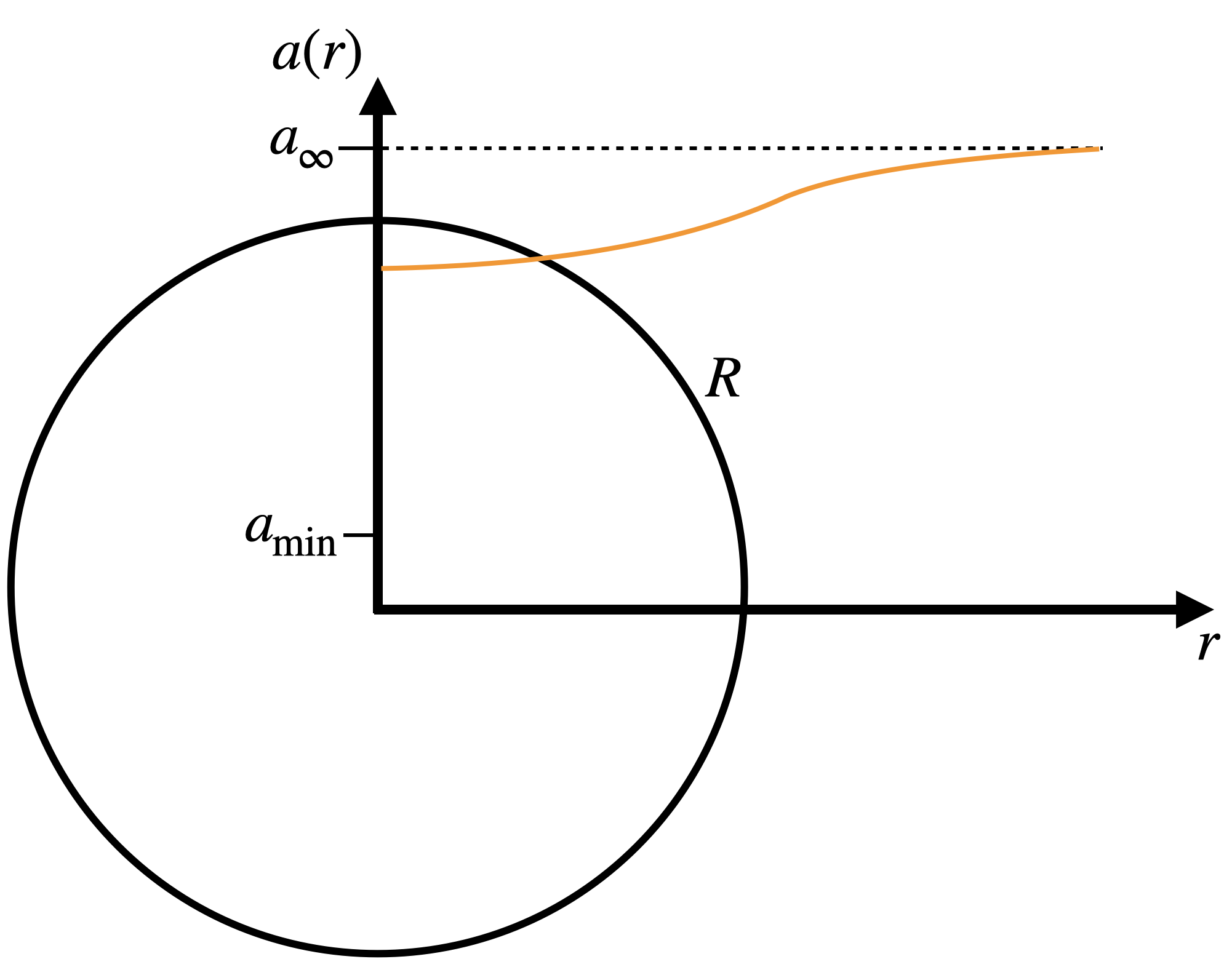}
    \caption{Behavior of the ALP inside astrophysical objects of radius $R$ when the effective potential is stabilized.~The asymptotic field value is $a_\infty$ but the density-dependent effective potential causes the field to deviate from this inside astrophysical bodies.~\textit{Left:}~Case where the field minimizes $V_{\rm eff}$.~There is a large gradient energy cost to doing this.~\textit{Right:}~If it is not energetically favorable for the field to minimize $V_{\rm eff}$ due to the gradient energy cost then the field does not make large excursions from $a_\infty$.~The field mediates a fifth-force proportional to $\nabla a$ so the  scenario in the left figure is expected to alter stellar structure.~If the density becomes sufficiently high that the ALP effective potential is destabilized then the field is expected to run away to large values (scenario not depicted).}
    \label{fig:astro_sketches}
\end{figure}

Figure \ref{fig:astro_sketches} shows the expected behavior of the ALP inside stars.~The CP-violating coupling causes the ALP VEV to be density-dependent, implying that the VEV at the center of  stars will differ from the asymptotic VEV at large distances.~The latter is set by minimizing the effective potential in the ambient background, which is the density in the host galaxy.~The former depends upon whether the minima exist, and, if so, whether it is energetically favorable for the field to minimize $V_{\rm eff}$ \cite{Burrage:2016bwy,Burrage:2017qrf}.~These inhomogeneous boundary conditions imply that stars will source an ALP field, giving rise to an ALP-mediated fifth-force between its constituents and, consequentially, altering its structure.~This raises two questions:~(1) what happens if $\rho>\rho_c$ at some radius within the object?~and (2) which objects, if any, can be used to constrain the CP-violating coupling?~We now address each of these in turn.

\subsection{ALP Destabilization Inside Stars}

\begin{figure}
    \centering \includegraphics[width=\textwidth]{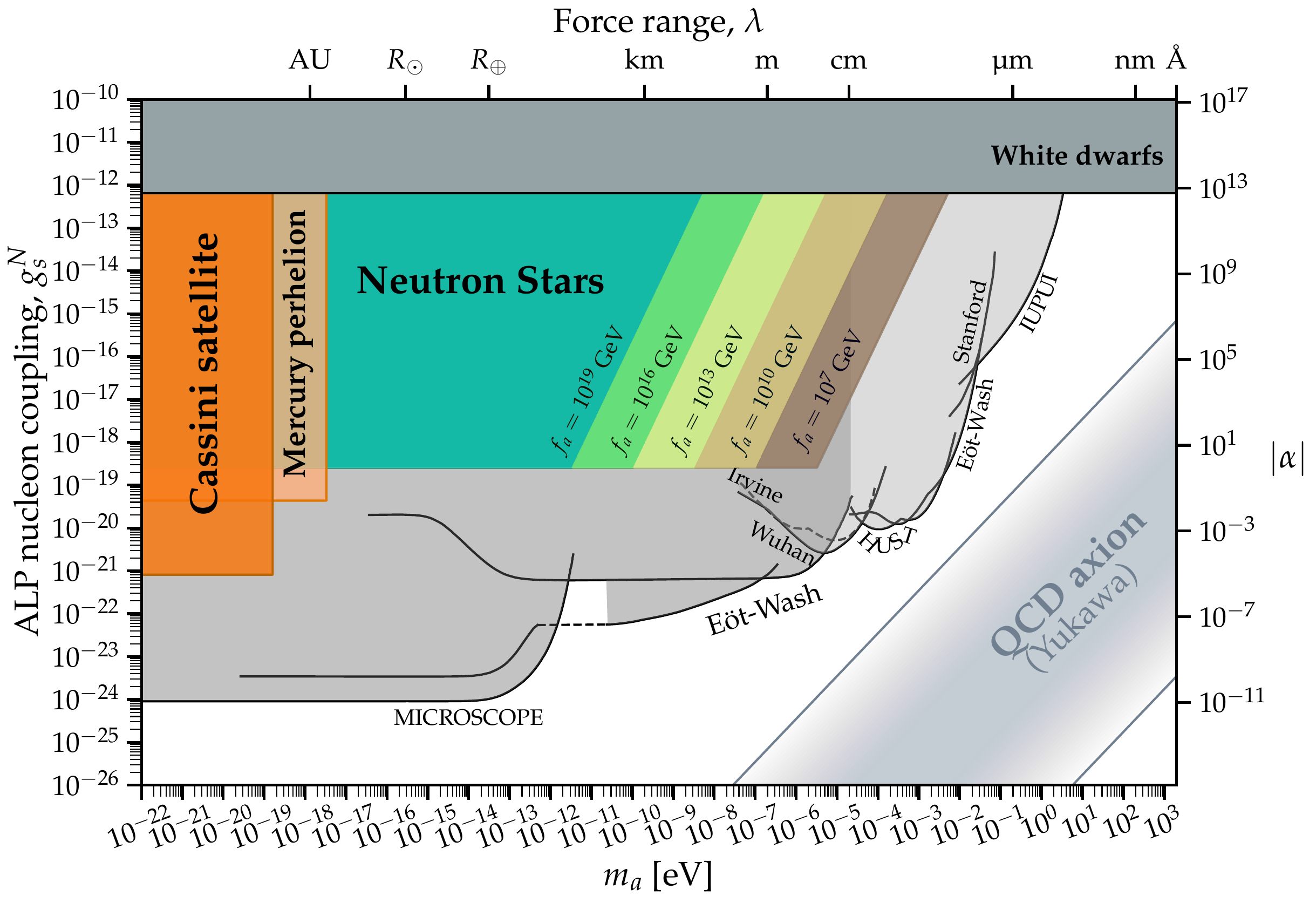}
    \caption{Region plot showing where the ALP potential becomes destabilized inside neutron stars, indicated  by $\rho_{NS} > \rho_c$, with a typical neutron star density of $\rho_{NS} = 10^{14}$ g/cm$^3$.~We show five regions, labeled by their respective axion decay constants, for which the change in the neutron mass detailed in Eq.~\eqref{eq:Delta_mNeutron_NS} is $\left\vert\Delta m_n/\bar{m}_n\right\vert \ge 1$.~Moreover, in yellow, we show the bounds from the perihelion of Mercury~\cite{Will:2014kxa}, and in red, the bounds from the Cassini satellite~\cite{Bertotti:2003rm};~both are derived in section~\ref{sec:SST}.~In grey, we show existing constraints from tests of the inverse-square law~\cite{Chen:2014oda,Lee:2020zjt,Kapner:2006si,Tan:2020vpf,Hoskins:1985tn}, white dwarf cooling~\cite{Bottaro:2023gep}, and WEP violation searches~\cite{Smith:1999cr, Berge:2017ovy}.~Lastly, the QCD axion region indicates where the QCD axion is expected to live~\cite{OHare:2020wah}.~The plot was generated by adapting the open-source code from O'Hare et al.~\cite{OHare:2020wah}.~}
    \label{fig:ALP_NS_bounds}
\end{figure}

If $\rho>\rho_c$ is reached at some radius within the star then the ALP effective potential is destabilized.~The potential energy is unbounded from below, and can in principle be reduced \textit{ad infinitium} by moving the ALP to increasingly negative values.~In practice, the ALP will not move to $-\infty$ because it is not the potential energy that must be minimized but the total energy.~Far from the object, the minimum energy configuration is one where the ALP minimizes $V_{\rm eff}$ in the ambient density, so there is a gradient energy cost to be paid for moving the field to negative values inside the star in order for the field to asymptote to the requisite value at large distances (see Fig.~\ref{fig:astro_sketches}).~We then expect that the field will move by a distance $\Delta a$ from it's asymptotic value $a_\infty$ such that
\begin{equation}
    \label{eq:E_General}
    E=\int \dd^3 \vec{x} \left(\frac12\dot{a}^2+\frac12|\nabla a|^2+V_{\rm eff}(a)\right)
\end{equation}
is minimized.~Assuming a static, spherically-symmetric object we can write $a(\vec{x})=a_\infty + \Delta a$, and make the further approximation that $\rho a(\vec{x})/\mu\gg V(a)$ appropriate for the destabilized regime.~If the star has a mass $M$ and radius $R$ we can approximate $\dd^3\vec{x}\sim R^3$, $\rho\sim 3M/4\pi R^3$, and $|\nabla a|\sim \Delta a/R$ to find 
\begin{equation}
    \label{eq:E_approximated}
    E\approx\frac12 R (\Delta a)^2+\frac{3M}{4\mu}(a_\infty+\Delta a).
\end{equation}
This is minimized when 
\begin{equation}
    \label{eq:fieldDestabilizedStars}
    \Delta a = - \frac{3 M}{4\mu R}.
\end{equation}
Equation~\eqref{eq:fieldDestabilizedStars} reveals that the largest field excursions happen in the most compact objects, which are neutron stars.~Following Eq.~\eqref{eq:nucleon_mass}, this implies that the neutron mass at the center of these stars differs from the laboratory measured value of then neutron mass by an amount
\begin{equation}
    \label{eq:neutronMassStars}
    \frac{\Delta m_n}{\bar{m}_n} =\frac{\Delta a}{\mu}=-\frac{3 M}{4 R\mu^2}.~
\end{equation}
For a typical neutron star with $M=1{\rm M}_\odot$ and $R=12$ km one has
\begin{equation}
    \label{eq:Delta_mNeutron_NS}
    \left\vert\frac{\Delta m_n}{\bar{m}_n}\right\vert=\left(\frac{3.7\times10^{18}\textrm{ GeV}}{\mu}\right)^2,
\end{equation}
implying that CP-violating couplings at the Planck-scale or smaller are problematic for values of $m_a$ and $f_a$ for which $\rho_{\rm NS}>\rho_c$.~We show this region in Fig.~\ref{fig:ALP_NS_bounds} for a typical neutron stars density $\rho_{\rm NS}=10^{14}$ g/cm$^3$.~In this region, the gravitational fine-structure constant $\alpha_{G}= Gm_n^2\rightarrow0$, implying that the neutron star will be unbound.

\subsection{Modified Stellar Structure}
\label{sec:I_dont_like_this_section}

We can gain some intuition for how and when stars are modified in the presence of the ALP CP-violating coupling by considering a static, spherically-symmetric non-relativistic object, for which equation~\eqref{eq:fiedlEOM_General} becomes 
\begin{equation}
    \frac{1}{r^2}\frac{\dd }{\dd r}\left(r^2\frac{\dd a(r)}{\dd r}\right)-\frac{\dd V_{\rm eff}(a)}{\dd a}=0.
\end{equation}
As discussed above, at large distances the field must tend to the asymptotic value $a_\infty$, which minimizes $V_{\rm eff}$ in the host galaxy's ambient density;~and the field  at the center will be displaced from this (see Fig.~\ref{fig:astro_sketches}).~In the case where $\rho>\rho_c$ in the core of the object the potential is destabilized, the field is displaced by an amount of order $3M/4\mu R$ (see Eq.~\eqref{eq:fieldDestabilizedStars}).~In the case where the potential is not destabilized energy considerations again dictate the central value.~Either the field will minimize the effective potential at the stellar central density if it is energetically favorable to do so, or a compromise between potential and gradient energy will be reached that minimizes the total energy.~In some theories, this behavior can give rise to novel phenomena such as  \textit{screening mechanisms} \cite{Joyce:2014kja,Burrage:2016bwy,Burrage:2017qrf,Baker:2019gxo,Sakstein:2020axg,Reyes:2024lzu} or \textit{spontaneous scalarization} \cite{Silva:2017uqg,Doneva:2022ewd}.~However, we do not expect these effects to arise in our theory because both the effective ALP mass and ALP-nucleon coupling do not vary significantly with density.~The coupling is constant and Eq.~\eqref{eq:meff} demonstrates that except in the limit $\rho \sim \rho_c$, the effective mass is always of order the bare mass $m_a$.~Thus, we expect that the field sourced by an object with mass $M$ will scale schematically as
\begin{equation}
    \label{eq:schematicFifthForce}
    a(r)\sim\frac{M}{4\pi\mu r}e^{-m_ar},
\end{equation}
demonstrating that deviations from general relativity (GR) will only manifest for parameters where the Compton wavelength $\lambda_{\rm C}\sim m_a^{-1}$ is larger than the length-scale of the object and 
\begin{equation}
    \label{eq:fifthForceRatio}
    \frac{F_{5}}{F_{\rm N}}=\frac{1}{4\pi G \mu^2}\gtrsim 1
\end{equation}
where $F_{5}=|\nabla a(r)|/\mu$ is the ALP-mediated fifth-force per unit mass \cite{Sakstein:2014jrq,Burrage:2016bwy,Burrage:2017qrf} and $F_{\rm N}$ is the gravitational acceleration.~Celestial bodies have length scales $R\gtrsim$ km, so, examining Fig.~\eqref{fig:ALP_NS_bounds}, the requirement in Eq.~\eqref{eq:fifthForceRatio} implies that any bounds will likely complement those coming from tests of the WEP.

\subsection{Neutron Star Structure} 

As discussed above, we expect the largest effects of ALPs, if any, inside neutron stars.~Equation \eqref{eq:Delta_mNeutron_NS} gives a new approximate bound on $\mu\gtrsim3.7\times10^{18}$ GeV for parameters $\{m_a,\, f_a,\,\mu \}$ where $\rho>\rho_c$ is obtained at some radius inside the star (see Fig~\ref{fig:ALP_NS_bounds}).~The precise bound is sensitive to the equation of state of nuclear matter of course.~This bound can equivalently be thought of as a bound on the regime of validity of the effective field theory (EFT).~The Yukawa coupling in \eqref{eq:ALP_Lagrangian_simple} scales as $a/\mu$ so we require $a/\mu\ll1$ for the theory to constitute a healthy EFT.~The neutron mass scales as $m_n(1+a/\mu)$ so equation Eq.~\eqref{eq:Delta_mNeutron_NS} demonstrates that, as expected, problems such as zero neutron mass arise precisely at energy scales of order the EFT cut-off $\mu$.~In order to avoid this, we must demand that $\mu\gtrsim3.7\times10^{18}$ GeV to avoid exiting the regime of validity of the EFT inside neutron stars.

Moving away from the extreme case above, we now turn our attention to couplings that satisfy the bound $\mu\gtrsim3.7\times10^{18}$ GeV and discuss whether the ALP CP-violating coupling can alter the structure of neutron stars.~The bound is already higher than the Planck scale, implying that we may be in the regime of quantum gravity.~Nonetheless, such couplings are often considered in the context of modified gravity.~Following our discussion in Sec.~\ref{sec:I_dont_like_this_section} above, we expect deviations from GR when $m_a\lesssim(10\textrm{ km}^{-1})\sim 10^{-11}$ eV and $\mu\lesssim\sqrt{2}M_{\rm Pl}\sim 3.4\times10^{18}$ GeV.~This is incompatible with the bound derived above, suggesting that deviations from GR will be negligible.~One caveat to this is that the bound does not apply if $f_a$ and $m_a$ are such that $\rho>\rho_c$.~We explored this possibility finding that deviations from GR are negligible.~In addition, there are further caveats that the bound on $\mu$ derived above made a series of approximations, and that Eq.~\eqref{eq:schematicFifthForce} and \eqref{eq:fifthForceRatio} were derived in the non-relativistic limit, which does not apply to neutron stars.~To eliminate these sources of uncertainty, we derived the modified Tolman-Oppenheimer-Volkovff (TOV) system of equations for CP-violating axions and solved them numerically for a realistic nuclear equation of state, APR \cite{Akmal:1998cf}, to find the modified mass-radius relation.~The equations and a description of our numerical code are given in Appendix \ref{sec:Omar'sMiricleCode}.~We explored the parameter range $\mu\in [10^{21},10^{25}]$ GeV, $f_a\in [10^{15},6\times10^{19}]$ GeV, $m_a\in [10^{-16},10^{-11}]$ eV.~We found negligible differences from the GR solutions, with deviations in the maximum masses being of order $\Delta M=10^{-3}{\rm M}_{\odot}$.~We therefore conclude that the new bound on $\mu$ that we have obtained is the only constraint that can be extracted from neutron stars.

\subsection{Solar System Tests}
\label{sec:SST}

As noted above, axions with masses smaller than the Compton wavelength of the object under consideration behave as massless scalar-tensor theories, which are constrained by the parameterized post-Newtonian (PPN) framework for testing GR in the solar system \cite{Will:2014kxa}.~In this subsection, we translate these bounds into bounds on $\mu$ as a function of $m_a$.~The two relevant parameters are $\gamma$, which controls the magnitude of the Shapiro time-delay effect and the angle of light bending by massive bodies;~and $\beta$, which determines the perihelion advance of orbiting bodies.~While enroute to Saturn in 2002, the Cassini satellite obtained conjunction with Earth, enabling a test of the Shapiro-time delay effect at a distance of $8.43$ Au \cite{Bertotti:2003rm}.~This constrained the parameter $\gamma$ to satisfy $|\gamma-1|<2\times10^{-5}$, which, in our case, applies when $m_a\lesssim  (8.43\textrm{ Au})^{-1}\sim 1.6\times10^{-19}$eV.~The parameter $\beta$ is constrained to satisfy $|\beta-1|<8\times10^{-5}$ using observations of the perihelion of Mercury \cite{Will:2014kxa}.~In our case, this bound applies for masses $m_a\lesssim R_{\rm Mercury}^{-1}\sim (0.4 \textrm{ Au})^{-1}\sim 3.3\times10^{-18}$ eV.~We can calculate $\gamma$ and $\beta$ for CP-violating ALPs when the bounds above apply by assuming the massless limit.~In this limit, one has \cite{Damour:1992we,Esposito-Farese:2004azw,Sakstein:2017pqi}
\begin{equation}
    \gamma-1=-4\frac{M_{\rm Pl}^2}{2M_{\rm Pl}^2+\mu^2}\textrm{ and }  \beta-1=-2\frac{M_{\rm pl}^4}{(2M_{\rm Pl}^2+\mu^2)^2}.
\end{equation}
The resultant bounds on $\mu$ are shown in Fig.~\ref{fig:ALP_NS_bounds}.~Quantitatively, the bound on $\gamma$ yields $\mu>1.16\times10^{21}$ GeV ($g_s^N <8.07\times10^{-22}$) and the bound on $\beta$ yields $\mu>2.16\times10^{19}$ GeV ($g_s^N <4.36\times10^{-20}$).

\section{Discussion and Conclusions}
\label{sec:discussion}

In this work we have explored the phenomenological consequences of Yukawa couplings between axion-like particles and nucleons.~Such couplings, which are induced by CP-violation in the UV, are expected to arise in generic models.~We identified a novel phenomenon whereby the ALP effective potential can be destabilized in sufficiently high nucleon densities larger than a model-dependent critical density.~In this regime, the field rolls over potentially large distances.~Below the critical density, the ALP vacuum expectation value is shifted, reducing its mass.

We explored the cosmology of CP-violating ALPs and found three distinct late-time regimes:
\begin{enumerate}
    \item ALPs with masses $m_a>10^{-32}$ eV behave like (a subfraction of) cold DM, similarly to the standard scenario.
    \item ALPs with masses in the range $m_a\sim 10^{-31}$--$10^{-32}$ eV execute oscillations about a time-dependent minimum.~The oscillating component behaves like cold DM while the component that tracks the minimum provides a source of dynamical dark energy with equation of state $w$ that exhibits detectable deviations from that of a cosmological constant, $w=-1$.~ALPs with masses in this range thus provide a unified description DM and dark energy.
    \item ALPs with masses $m_a\sim 10^{-33}$ eV act as dynamical dark energy with an equation of state that deviates from that of a cosmological constant.
\end{enumerate}
At early times, the ALP potential is destabilized and can roll over large distances.~The specific time depends upon the model parameters.~This rolling is problematic because the nucleon masses would change by a factor $\Delta m_N=\Delta a/\mu$.~Requiring that any such change between big bang nucelosynthesis and the CMB is smaller than $10\%$ imposes the bound $\mu\gtrsim5\times10^{18}$ GeV.

We also explored the effects of ALP CP-violating couplings for neutron stars.~We found that the large densities inside neutron stars can be sufficient to destabilize the ALP if $\mu\gtrsim3.7\times10^{18}$ GeV.~When this condition is imposed and the potential is stabilized, the effects of axion-mediated Yuakwa forces are negligible.~Finally, we derived bounds on the model parameters imposed by solar system tests of post-Newtonian gravity.

Our preliminary study lays the foundation for follow-up investigations along several lines.~One possibility would be to build and constrain models of ALP dark energy.~Another would be to build models of ALP early dark energy aimed at resolving the Hubble tension.~The canonical early dark energy model utilizes ALP-like periodic potentials \cite{Poulin:2018cxd,Poulin:2023lkg} but it faces theoretical challenges \cite{Rudelius:2022gyu,Ramadan:2023ivw}.~The model parameters are tuned such that the onset of early dark energy coincides with matter-radiation equality, and the leading-order ALP potential is fine-tuned to zero to avoid predicting too much DM at late times.~It would be interesting to explore the possibility of using a combination of the time-varying axion mass and destabilization as building blocks for a more natural model.

Another application would be to relaxion models, which may resolve the Higgs or cosmological constant fine-tuning problems \cite{Abbott:1984qf,Graham:2015cka}.~These utilize a similar potential to ours in the destabilized regime, with the notable  difference that the destabilization is due to a linear term in the bare potential so is ever-present.~The field is ultimately stabilized by ALP-gauge field strength couplings.~Including CP-violating Yukawa couplings to matter would provide an alternative method for stabilizing the relaxion at late times.

Finally, the novel phenomenological features we have identified could be imported into more general theories e.g., dark sector ALPs coupled to dark gauge groups.~The parameters in such theories would be subject to fewer observational bounds, enlarging the space for model building.

\section*{Software}

Mathematica v13.2.1.0, Python v3.10.14, SciPy v1.13.0, open-source plotting code from \href{https://github.com/cajohare/AxionLimits/blob/master/AxionCPV.ipynb}{O'Hare et al.~(2020)}~\cite{OHare:2020wah}.

\section*{Acknowledgements}

We are grateful for discussions with Philippe Brax, Cliff Burgess, Benjamin Elder, and Samuel D.~McDermott.~This material is based upon work supported by the National Science Foundation under Grant No.~2207880.~DC is supported by the STFC under Grant No.~ST/T001011/1.~The technical support and advanced computing resources from University of Hawaii Information Technology Services – Cyberinfrastructure, funded in part by the National Science Foundation CC* awards $\#2201428$ and $\#2232862$ are gratefully acknowledged.

\appendix

\section{Equivalence of Yukawa couplings and Conformal Scalar-Tensor Theories}
\label{app:MG}

In this Appendix, we review the equivalence between conformal scalar-tensor theories of gravity and Yukawa couplings, and use this equivalence to derive some results used in the main text.~

Our starting point is a conformal scalar-tensor theory in the \textit{Einstein frame}.~This theory contains a graviton described by the \textit{Einstein frame metric} $g_{\mu\nu}$, a scalar $a$ (the ALP), and an arbitrary number of matter fermions ${\Psi}_j$.~These are coupled to the conformally-rescaled \textit{Jordan frame metrics} $G^{(j)}_{\mu\nu}=A^2_j(a)g_{\mu\nu}$
\begin{equation}
    \label{eq:action_conformalST_general}
    S=\int \dd^4 x\sqrt{-g}\left[ \frac{R(g)}{16\pi G}-\frac12\nabla_\mu a\nabla^\mu a-V(a)\right]+S_m[G^{(j)}_{\mu\nu};{\Psi}_j],
\end{equation}
where $A(\phi)$  are free \textit{coupling functions}.~The final term indicates that all indices in the action for the matter fermions are contracted with their own Jordan frame metric $G^{(j)}$.~It is this coupling that gives rise to modifications of general relativity.~The fermions move on geodesics of $G^{(j)}_{\mu\nu}$ whereas it is $g_{\mu\nu}$ that satisfies the Einstein equations.~In this description, the space-time curvature sourced by matter differs from the space-time governing the motion of particles, and one can view each particle as living in its own distinct spacetime $G^{(j)}_{\mu\nu}$.

To see that this coupling is equivalent to a Yukawa coupling, we expand out the action for fermions as
\begin{equation}
\label{eq:action_conformalST_specific_fermions}
    S_m[G^{(j)}_{\mu\nu};{\Psi}_j]=\sum_j\int\dd^4 x\sqrt{-G^{(j)}}\left[i\bar{{\Psi}}_jE^\mu_{(j)a}\gamma^a\overset{\text{\tiny$\bm\leftrightarrow$}}{\partial}_\mu{\Psi}_j-m_j\bar{\Psi}_j\Psi_j\right],
\end{equation}
where $E^a_{(j)\mu}$ is the vierbein for $G^{(j)}$ i.e., $G_{\mu\nu}^{(j)} {E}^\mu_{(j)a}{E}_{(j)b}^\nu=\eta_{ab}$, $\{\gamma^a,\gamma^b\}=\eta^{ab}$, and $\overset{\text{\tiny$\bm\leftrightarrow$}}{\partial}_\mu=(\overset{\text{\tiny$\bm\rightarrow$}}{\partial}_\mu-\overset{\text{\tiny$\bm\leftarrow$}}{\partial}_\mu)/2$;~this formalism has been used to avoid the need to include the spin-connection.~Expanding this action in terms of the Einstein frame metric using $\sqrt{-G^{(j)}}=A^4_j(a)\sqrt{-g}$, $E_{(j)a}^\mu=A_j^{-1}(a)e_a^\mu$ with $e_a^\mu$ the Einstein frame vierbein i.e., $g_{\mu\nu}e_a^\mu e_b^\nu=\eta_{ab}$ one finds
\begin{equation}
    \label{eq:action_EF_non_normalized}
    S_m[g_{\mu\nu};{\Psi}_j]=\sum_j\int\dd^4 x\sqrt{-g}\left[i A_j^3(a)\bar{{\Psi}}e^\mu_{a}\gamma^a\overset{\text{\tiny$\bm\leftrightarrow$}}{\partial}_\mu{\Psi}-m_jA_j^4(a)\bar{\Psi}_j\Psi_j\right].
\end{equation}
One can then canonically normalize the fermion fields as $\psi_j=A^{-\frac32}(a)\Psi_j$ to find
\begin{equation}
    \label{eq:action_EF_normalized}
    S_m[g_{\mu\nu};{\psi}_j]=\sum_j\int\dd^4 x\sqrt{-g}\left[i\bar{{\psi}}e^\mu_{a}\gamma^a\overset{\text{\tiny$\bm\leftrightarrow$}}{\partial}_\mu{\psi}-m_jA_j(a)\bar{\psi}_j\psi_j\right].
\end{equation}
A Yukawa coupling of $a$ to the field $\psi_j$ is then seen to be equivalent to
\begin{equation}
    A_j(a)=1+\frac{a}{\mu_j} = 1+g_j\frac{a}{m_j}
\end{equation}
where the mass scales $\mu_j$ are introduced on grounds of dimension and the second expression introduces the dimensionless Yukawa coupling $g_j=m_j/\mu_j$.

The equation of motion for the scalar can be derived directly from the action~\eqref{eq:action_conformalST_general}.~Doing so, one has
\begin{equation}
    \label{eq:EOM_derivation1}
    \Box a - \frac{\dd V(a)}{\dd a} + \frac{1}{\sqrt{-g}} \frac{\delta S_m}{\delta a} =  0.
\end{equation}
The final term can be expanded as
\begin{equation}
    \label{eq:EOM_derivation2}
    \frac{1}{\sqrt{-g}} \frac{\delta S}{\delta a} = \sum_j \frac{1}{\sqrt{-g}} \frac{\delta S}{\delta G_{\mu\nu}^{(j)}}\frac{\delta G_{\mu\nu}^{(j)}}{\delta a}=\sum_j  A_j^4(a) G_{\mu\nu}^{(j)}\Theta_j^{\mu\nu}  \frac{\dd \ln A_j(a)}{\dd a},
\end{equation}
where $\Theta_j^{\mu\nu}=\frac{2}{\sqrt{-G^{(j)}}}\frac{\delta S}{\delta G_{\mu\nu}^{(j)}}$ is the trace of the fermion energy-momentum tensor for $\Psi_j$.~We wish to work with the canonically-normalized field $\psi_j$ in the Einstein frame.~Making this change, one has  $G_{\mu\nu}^{(j)}\Theta_j^{\mu\nu}=A^{-4}_jg_{\mu\nu}T_j^{\mu\nu}=A_j^{-4}T_j$ where $T_j^{\mu\nu}$ is the energy-momentum tensor for $\psi_j$ derived from the action \eqref{eq:action_EF_normalized} and $T_j$ is its trace \cite{Sakstein:2014jrq}.~The EOM in equation \eqref{eq:EOM_derivation1} is then
\begin{equation}
    \label{eq:EOM_derivation3}
    \Box a - \frac{\dd V(a)}{\dd a} + \sum_j \frac{\dd\ln A_j(a)}{\dd a}T_j =  0,
\end{equation}
showing that the dynamics of the scalar are governed by an effective potential $V_{\rm eff}=V(a)-\sum_jT_j\ln(A_j)$, which is equivalent to Eq.~\eqref{eq:fiedlEOM_General} after coupling to nucleons with equal strength and working to first-order in $a/\mu$.

\section{Stellar Structure Equations in the Presence of CP-Violating ALPs}
\label{sec:Omar'sMiricleCode}

In this Appendix, we derive the Tolman-Oppenheimer-Volkov equations governing the structure of static spherically symmetric neutron stars in the presence of ALP-nucleon couplings, and explain our numerical procedure for solving them.~We begin by deriving the modified TOV equations.~The equations to be solved are the Einstein and Klein-Gordon equations.~We work in the Einstein frame and in the limit where $a/\mu\ll1$ so that factors of $A(a)$ (but not $A'(a)$) can be neglected.~We use the standard gauge for the metric
\begin{equation}\label{eq:star_metic}
     \mathrm{d}s^2 = -e^{\nu(r)} dt^2 + e^{\lambda(r)} dr^2 + r^2 d\Omega^2,
\end{equation}
where $\dd\Omega^2$ is the line-element on a unit 2-sphere;~and make the ansatz $a=a(r)$, mandated by the symmetries of the system.~We model the star as a perfect fluid with pressure $p$ and density $\rho$ with corresponding energy-momentum tensor $T^{\mu\nu}=(\rho + p) u^\mu u^\nu + p g^{\mu\nu}$ with $u^\mu=(-e^{-\frac{\nu}{2}},0,0,0)$ the four velocity in this gauge.~Inside the star, these ansatzes leads to the following system of equations 
\begin{gather} 
    \frac{dp}{dr} = - 
    \frac{1}{2} \left(\rho + p\right)\frac{d\nu}{dr} + 
    \left(3 p - \rho\right) \frac{f_a}{\mu}  \frac{d\theta}{dr} \label{eq:modified_TOV},\\
    \frac{d\nu}{dr} = \frac{e^{\lambda}-1}{r} + \left[8\pi r G \biggl(p - \rho \frac{f_a }{\mu}- m_a^2 f_a^2 (1-\cos{(\theta)}) \biggr) e^\lambda+4\pi r G f_a^2 \left(\frac{d\theta}{dr}\right)^2\right]\label{eq:inside_metric_pot},\\
    \begin{split}
    \frac{d\lambda}{dr} = \frac{1-e^{\lambda}}{r} + \Biggl[8\pi r G \biggl(\rho \left(1+\frac{f_a}{\mu}\right)+ &m_a^2 f_a^2 \left(1-\cos{(\theta)}\right)\biggr)  e^\lambda+4\pi r G f_a^2 \left(\frac{d\theta}{dr}\right)^2\Biggr]\label{eq:inside_lambda_pot},
    \end{split}\\
    \frac{d^2 \theta}{dr^2} + \frac{d\theta}{dr}\left(\frac{2}{r}-\frac{1}{2}\frac{d\lambda}{dr}+\frac{1}{2}\frac{d\nu}{dr}\right) -  m_a^2 e^{\lambda} \sin{(\theta)} = \frac{\rho  e^\lambda}{f_a\mu}\label{eq:inside_KG},
\end{gather}
where $\theta \equiv a/f_a$.~We use the APR equation of state~\cite{Akmal:1998cf} to close our system of equations, enabling us to express $p\equiv p(\rho)$.

The system above is an initial-boundary value problem --- the equations for the metric potentials and the pressure/density are initial value problems while the Klein-Gordon equation for the scalar is a boundary value problem.~The central conditions are $\rho(0)=\rho_{\rm central}$, which gives $p_{\rm central}$ via the EOS;~$\lambda(0)=0$, reflecting the fact that there is no mass at $r=0$;~$\nu(0)=\nu_{\rm central}$;~and $\theta'(0)=0$, which is mandated by spherical symmetry.~Physically, $\rho_{\rm central}$ is a free parameter that  determines the star's mass and radius, while $\nu_{\rm central}$ is a choice of time-coordinate/units that is shifted as $\nu_{\rm central}\rightarrow\nu_{\rm central}+\nu_0$ under the gauge transformation $t\rightarrow e^{\nu_0/2}t$.~We will make use of this  to extract the star's mass.~The central ALP value $\theta(0)=\theta_{\rm central}$ is fixed by the requirement that $\lim_{r\rightarrow\infty} \theta(r)=\theta_\infty$ where $\theta_\infty=a_\infty/f_a$ is the asymptotic field value, which minimizes the effective potential in the ambient density $\rho_\infty$ surrounding the neutron star.~In what follows we will take $\theta_\infty=0$ corresponding to  $\rho_\infty=0$.~Strictly speaking, the physical ambient density is the galactic density, $\rho_{\rm gal}\sim 10^{-24}$ g/cm$^3$ but, practically, this shifts the minimum from $a_\infty=0$ by a negligible amount.~The radius of the star $R$ corresponds to the point where $\rho(R)=0$.~The scalar, the metric potentials, and their derivatives are continuous across this boundary.

Due to the singularity at $r=0$, the numerical integration  cannot begin at the star's center.~Instead, we define the initial conditions at a small distance from the center, $r_i$, where $r_i\ll R$ with $R$ being the radius of a neutron star, typically on the order of $10^6$ cm.~We choose $r_i = 10^{-15}$ cm.~For consistency, we derive the initial conditions for metric potentials, field, and density at $r=r_i$ by expanding them to second-order in $r$ as
\begin{align}
    \rho_i &= \rho_{\rm central} + \alpha r_i^2,\label{eq:rho_init} \\
    \nu_i &= \nu_{\rm central} + \beta r_i^2,\label{eq:nu_init} \\
    \lambda_i &= \gamma r_i^2,\label{eq:lambda_init} \\
    \theta_i &= \theta_{\rm central} + \eta r_i^2,\label{eq:theta_init}
\end{align}
and using the modified TOV equations to find the unknown coefficients $\{\alpha,\beta,\lambda,\eta\}$ as functions of the central values and ALP parameters.~Notably, $\lambda_{\rm central}$ is zero, reflecting the fact  that no mass in enclosed at $r=0$ is zero, as anticipated.~The final expressions are cumbersome so we do not give them here.~We imposed the asymptotic conditions at a final radius $r_f=2000$ km.

We solved the TOV system above using a custom algorithm, which will be released upon publication of this work: 
\begin{enumerate}
    \item The \texttt{LSODA} method~\cite{Hindmarsh1983ODEPACK,2019ascl.soft05021H} was first utilized to solve the interior system as an initial value problem with an initial guess for the central field value $\theta(0)=\theta_{\rm central}$ found by using  Eq.~\eqref{eq:fieldDestabilizedStars}  assuming a constant density star with $\rho=\rho_{\rm cent}$ to find the field value that minimizes the Hamiltonian: 
\begin{equation}\label{eq:fieldStarIC}
    \theta_{\rm central} = -\frac{\pi R^2\rho_{\rm cent}}{\mu f_a}.
\end{equation}
The system was integrated until the radius $R$ where $\rho(R)=0$, yielding a value for $\theta(R)=\theta_-$.
    \item  Next, the \texttt{solve\_bvp} function from \texttt{SciPy} was used to solve the exterior boundary value problem.~This method utilizes a fourth-order collocation algorithm with residual control~\cite{Kierzenka2001ABS}, and solves the collocation system through a damped Newton method with an affine-invariant criterion function~\cite{Ascher1995}.~We use $\theta(r_f)=0$ and $\theta'(R)$ obtained in the previous step as boundary conditions.~This step yields   $\theta(R)=\theta_+$ such that $\lim_{r\rightarrow}\infty=0$.
    \item The \texttt{Nedler-Mead} minimization algorithm \cite{Gao:2012guu} was employed to minimize a pre-defined \texttt{cost} function without the need for gradients, making it suitable for complicated problems where the gradients are computationally expensive or inaccessible.~We used a mean squared error (MSE) cost function ${ \rm cost} = |\theta'(r_f)-0|^2 + |\theta_- - \theta_+|^2$.~To restrict ourselves to physical solutions, we penalized solutions where $\theta(r)>0$ and null solutions $\theta(r<R) = 0$.~Due to the difference in the scales of each term, we normalize the terms with equal weights to ensure that minimizing the cost is fair to both terms.
\end{enumerate}
This algorithm yields numerical solutions for $\{\nu(r),\lambda(r),\theta(r),\rho(r)\}$.~To find the star's mass, we use the asymptotic expansion for the metric potential 
\begin{equation}
\label{eq:ns_mass_metric_pot}
    \lim_{r\rightarrow\infty}\nu(r)=A \left( 1-\frac{2GM}{r}\right)
\end{equation}
where $A$ is a constant that accounts for the arbitrariness of the time coordinate discussed above.~The values of $A$ and $M$ were extracted by matching the numerical solutions for $\nu(r)$ and $\nu'(r)$ onto this.

\newpage
\bibliographystyle{JHEP}
\bibliography{main.bib}

\providecommand{\href}[2]{#2}\begingroup\raggedright\begin{thebibliography}{10}

\bibitem{Svrcek:2006yi}
P.~Svrcek and E.~Witten, \emph{{Axions In String Theory}},
  \href{https://doi.org/10.1088/1126-6708/2006/06/051}{\emph{JHEP} {\bfseries
  06} (2006) 051} [\href{https://arxiv.org/abs/hep-th/0605206}{{\ttfamily
  hep-th/0605206}}].

\bibitem{Marsh:2015xka}
D.J.E.~Marsh, \emph{{Axion Cosmology}},
  \href{https://doi.org/10.1016/j.physrep.2016.06.005}{\emph{Phys. Rept.}
  {\bfseries 643} (2016) 1} [\href{https://arxiv.org/abs/1510.07633}{{\ttfamily
  1510.07633}}].

\bibitem{Chadha-Day:2021szb}
F.~Chadha-Day, J.~Ellis and D.J.E.~Marsh, \emph{{Axion dark matter: What is it
  and why now?}}, \href{https://doi.org/10.1126/sciadv.abj3618}{\emph{Sci.
  Adv.} {\bfseries 8} (2022) abj3618}
  [\href{https://arxiv.org/abs/2105.01406}{{\ttfamily 2105.01406}}].

\bibitem{Alexander:2023wgk}
S.~Alexander, H.~Gilmer, T.~Manton and E.~McDonough,
  \emph{{\ensuremath{\pi}-axion and \ensuremath{\pi}-axiverse of dark QCD}},
  \href{https://doi.org/10.1103/PhysRevD.108.123014}{\emph{Phys. Rev. D}
  {\bfseries 108} (2023) 123014}
  [\href{https://arxiv.org/abs/2304.11176}{{\ttfamily 2304.11176}}].

\bibitem{Apers:2024ffe}
F.~Apers, J.P.~Conlon, E.J.~Copeland, M.~Mosny and F.~Revello, \emph{{String
  Theory and the First Half of the Universe}},
  \href{https://arxiv.org/abs/2401.04064}{{\ttfamily 2401.04064}}.

\bibitem{Peccei:1977hh}
R.D.~Peccei and H.R.~Quinn, \emph{{CP Conservation in the Presence of
  Instantons}}, \href{https://doi.org/10.1103/PhysRevLett.38.1440}{\emph{Phys.
  Rev. Lett.} {\bfseries 38} (1977) 1440}.

\bibitem{Peccei:1977ur}
R.D.~Peccei and H.R.~Quinn, \emph{{Constraints Imposed by CP Conservation in
  the Presence of Instantons}},
  \href{https://doi.org/10.1103/PhysRevD.16.1791}{\emph{Phys. Rev. D}
  {\bfseries 16} (1977) 1791}.

\bibitem{Peccei:2006as}
R.D.~Peccei, \emph{{The Strong CP problem and axions}},
  \href{https://doi.org/10.1007/978-3-540-73518-2_1}{\emph{Lect. Notes Phys.}
  {\bfseries 741} (2008) 3}
  [\href{https://arxiv.org/abs/hep-ph/0607268}{{\ttfamily hep-ph/0607268}}].

\bibitem{Kim:2008hd}
J.E.~Kim and G.~Carosi, \emph{{Axions and the Strong CP Problem}},
  \href{https://doi.org/10.1103/RevModPhys.82.557}{\emph{Rev. Mod. Phys.}
  {\bfseries 82} (2010) 557} [\href{https://arxiv.org/abs/0807.3125}{{\ttfamily
  0807.3125}}].

\bibitem{OHare:2024nmr}
C.A.J.~O'Hare, \emph{{Cosmology of axion dark matter}},
  \href{https://doi.org/10.22323/1.454.0040}{\emph{PoS} {\bfseries
  COSMICWISPers} (2024) 040}
  [\href{https://arxiv.org/abs/2403.17697}{{\ttfamily 2403.17697}}].

\bibitem{Freese:1990rb}
K.~Freese, J.A.~Frieman and A.V.~Olinto, \emph{{Natural inflation with pseudo -
  Nambu-Goldstone bosons}},
  \href{https://doi.org/10.1103/PhysRevLett.65.3233}{\emph{Phys. Rev. Lett.}
  {\bfseries 65} (1990) 3233}.

\bibitem{Pajer:2013fsa}
E.~Pajer and M.~Peloso, \emph{{A review of Axion Inflation in the era of
  Planck}}, \href{https://doi.org/10.1088/0264-9381/30/21/214002}{\emph{Class.
  Quant. Grav.} {\bfseries 30} (2013) 214002}
  [\href{https://arxiv.org/abs/1305.3557}{{\ttfamily 1305.3557}}].

\bibitem{Croon:2014dma}
D.~Croon and V.~Sanz, \emph{{Saving Natural Inflation}},
  \href{https://doi.org/10.1088/1475-7516/2015/02/008}{\emph{JCAP} {\bfseries
  02} (2015) 008} [\href{https://arxiv.org/abs/1411.7809}{{\ttfamily
  1411.7809}}].

\bibitem{Croon:2015fza}
D.~Croon, V.~Sanz and J.~Setford, \emph{{Goldstone Inflation}},
  \href{https://doi.org/10.1007/JHEP10(2015)020}{\emph{JHEP} {\bfseries 10}
  (2015) 020} [\href{https://arxiv.org/abs/1503.08097}{{\ttfamily
  1503.08097}}].

\bibitem{Frieman:1995pm}
J.A.~Frieman, C.T.~Hill, A.~Stebbins and I.~Waga, \emph{{Cosmology with
  ultralight pseudo Nambu-Goldstone bosons}},
  \href{https://doi.org/10.1103/PhysRevLett.75.2077}{\emph{Phys. Rev. Lett.}
  {\bfseries 75} (1995) 2077}
  [\href{https://arxiv.org/abs/astro-ph/9505060}{{\ttfamily
  astro-ph/9505060}}].

\bibitem{Poulin:2018cxd}
V.~Poulin, T.L.~Smith, T.~Karwal and M.~Kamionkowski, \emph{{Early Dark Energy
  Can Resolve The Hubble Tension}},
  \href{https://doi.org/10.1103/PhysRevLett.122.221301}{\emph{Phys.~Rev.~Lett.}
  {\bfseries 122} (2019) 221301}
  [\href{https://arxiv.org/abs/1811.04083}{{\ttfamily 1811.04083}}].

\bibitem{McDonough:2022pku}
E.~McDonough and M.~Scalisi, \emph{{Towards Early Dark Energy in string
  theory}}, \href{https://doi.org/10.1007/JHEP10(2023)118}{\emph{JHEP}
  {\bfseries 10} (2023) 118}
  [\href{https://arxiv.org/abs/2209.00011}{{\ttfamily 2209.00011}}].

\bibitem{Baryakhtar:2022hbu}
M.~Baryakhtar et~al., \emph{{Dark Matter In Extreme Astrophysical
  Environments}},  in \emph{{Snowmass 2021}}, 3, 2022
  [\href{https://arxiv.org/abs/2203.07984}{{\ttfamily 2203.07984}}].

\bibitem{Adams:2022pbo}
C.B.~Adams et~al., \emph{{Axion Dark Matter}},  in \emph{{Snowmass 2021}}, 3,
  2022 [\href{https://arxiv.org/abs/2203.14923}{{\ttfamily 2203.14923}}].

\bibitem{DiLuzio:2020wdo}
L.~Di~Luzio, M.~Giannotti, E.~Nardi and L.~Visinelli, \emph{{The landscape of
  QCD axion models}},
  \href{https://doi.org/10.1016/j.physrep.2020.06.002}{\emph{Phys. Rept.}
  {\bfseries 870} (2020) 1} [\href{https://arxiv.org/abs/2003.01100}{{\ttfamily
  2003.01100}}].

\bibitem{Sikivie:2020zpn}
P.~Sikivie, \emph{{Invisible Axion Search Methods}},
  \href{https://doi.org/10.1103/RevModPhys.93.015004}{\emph{Rev. Mod. Phys.}
  {\bfseries 93} (2021) 015004}
  [\href{https://arxiv.org/abs/2003.02206}{{\ttfamily 2003.02206}}].

\bibitem{Moody:1984ba}
J.E.~Moody and F.~Wilczek, \emph{{NEW MACROSCOPIC FORCES?}},
  \href{https://doi.org/10.1103/PhysRevD.30.130}{\emph{Phys. Rev. D} {\bfseries
  30} (1984) 130}.

\bibitem{Georgi:1986kr}
H.~Georgi and L.~Randall, \emph{{Flavor Conserving CP Violation in Invisible
  Axion Models}},
  \href{https://doi.org/10.1016/0550-3213(86)90022-2}{\emph{Nucl. Phys. B}
  {\bfseries 276} (1986) 241}.

\bibitem{Bertolini:2020hjc}
S.~Bertolini, L.~Di~Luzio and F.~Nesti, \emph{{Axion-mediated forces, CP
  violation and left-right interactions}},
  \href{https://doi.org/10.1103/PhysRevLett.126.081801}{\emph{Phys. Rev. Lett.}
  {\bfseries 126} (2021) 081801}
  [\href{https://arxiv.org/abs/2006.12508}{{\ttfamily 2006.12508}}].

\bibitem{OHare:2020wah}
C.A.J.~O'Hare and E.~Vitagliano, \emph{{Cornering the axion with $CP$-violating
  interactions}},
  \href{https://doi.org/10.1103/PhysRevD.102.115026}{\emph{Phys. Rev. D}
  {\bfseries 102} (2020) 115026}
  [\href{https://arxiv.org/abs/2010.03889}{{\ttfamily 2010.03889}}].

\bibitem{DiLuzio:2021jfy}
L.~Di~Luzio, \emph{{CP-violating axions}},
  \href{https://doi.org/10.22323/1.398.0513}{\emph{PoS} {\bfseries EPS-HEP2021}
  (2022) 513} [\href{https://arxiv.org/abs/2108.09071}{{\ttfamily
  2108.09071}}].

\bibitem{Barbieri:1996vt}
R.~Barbieri, A.~Romanino and A.~Strumia, \emph{{On axion mediated macroscopic
  forces again}},
  \href{https://doi.org/10.1016/0370-2693(96)00957-4}{\emph{Phys. Lett. B}
  {\bfseries 387} (1996) 310}
  [\href{https://arxiv.org/abs/hep-ph/9605368}{{\ttfamily hep-ph/9605368}}].

\bibitem{Pospelov:1997uv}
M.~Pospelov, \emph{{CP odd interaction of axion with matter}},
  \href{https://doi.org/10.1103/PhysRevD.58.097703}{\emph{Phys. Rev. D}
  {\bfseries 58} (1998) 097703}
  [\href{https://arxiv.org/abs/hep-ph/9707431}{{\ttfamily hep-ph/9707431}}].

\bibitem{Coleman:1985rnk}
S.~Coleman, \emph{{Aspects of Symmetry}: {Selected Erice Lectures}}, Cambridge
  University Press, Cambridge, U.K. (1985),
  \href{https://doi.org/10.1017/CBO9780511565045}{10.1017/CBO9780511565045}.

\bibitem{Csaki:1998vv}
C.~Csaki and H.~Murayama, \emph{{Instantons in partially broken gauge groups}},
  \href{https://doi.org/10.1016/S0550-3213(98)00448-9}{\emph{Nucl. Phys. B}
  {\bfseries 532} (1998) 498}
  [\href{https://arxiv.org/abs/hep-th/9804061}{{\ttfamily hep-th/9804061}}].

\bibitem{Wetterich:2014bma}
C.~Wetterich, \emph{{Modified gravity and coupled quintessence}},
  \href{https://doi.org/10.1007/978-3-319-10070-8_3}{\emph{Lect. Notes Phys.}
  {\bfseries 892} (2015) 57} [\href{https://arxiv.org/abs/1402.5031}{{\ttfamily
  1402.5031}}].

\bibitem{Burrage:2018dvt}
C.~Burrage, E.J.~Copeland, P.~Millington and M.~Spannowsky, \emph{{Fifth
  forces, Higgs portals and broken scale invariance}},
  \href{https://doi.org/10.1088/1475-7516/2018/11/036}{\emph{JCAP} {\bfseries
  11} (2018) 036} [\href{https://arxiv.org/abs/1804.07180}{{\ttfamily
  1804.07180}}].

\bibitem{Sakstein:2013pda}
J.~Sakstein, \emph{{Stellar Oscillations in Modified Gravity}},
  \href{https://doi.org/10.1103/PhysRevD.88.124013}{\emph{Phys. Rev. D}
  {\bfseries 88} (2013) 124013}
  [\href{https://arxiv.org/abs/1309.0495}{{\ttfamily 1309.0495}}].

\bibitem{Sakstein:2014jrq}
J.~Sakstein, \emph{{Astrophysical Tests of Modified Gravity}}, Ph.D. thesis,
  Cambridge U., DAMTP, 2014.
\newblock \href{https://arxiv.org/abs/1502.04503}{{\ttfamily 1502.04503}}.
\newblock 10.17863/CAM.16133.

\bibitem{Burrage:2016bwy}
C.~Burrage and J.~Sakstein, \emph{{A Compendium of Chameleon Constraints}},
  \href{https://doi.org/10.1088/1475-7516/2016/11/045}{\emph{JCAP} {\bfseries
  11} (2016) 045} [\href{https://arxiv.org/abs/1609.01192}{{\ttfamily
  1609.01192}}].

\bibitem{Burrage:2017qrf}
C.~Burrage and J.~Sakstein, \emph{{Tests of Chameleon Gravity}},
  \href{https://doi.org/10.1007/s41114-018-0011-x}{\emph{Living Rev. Rel.}
  {\bfseries 21} (2018) 1} [\href{https://arxiv.org/abs/1709.09071}{{\ttfamily
  1709.09071}}].

\bibitem{Sakstein:2017pqi}
J.~Sakstein, \emph{{Tests of Gravity with Future Space-Based Experiments}},
  \href{https://doi.org/10.1103/PhysRevD.97.064028}{\emph{Phys. Rev. D}
  {\bfseries 97} (2018) 064028}
  [\href{https://arxiv.org/abs/1710.03156}{{\ttfamily 1710.03156}}].

\bibitem{Sakstein:2018fwz}
J.~Sakstein, \emph{{Astrophysical tests of screened modified gravity}},
  \href{https://doi.org/10.1142/S0218271818480085}{\emph{Int. J. Mod. Phys. D}
  {\bfseries 27} (2018) 1848008}
  [\href{https://arxiv.org/abs/2002.04194}{{\ttfamily 2002.04194}}].

\bibitem{Baker:2019gxo}
T.~Baker et~al., \emph{{Novel Probes Project: Tests of gravity on astrophysical
  scales}}, \href{https://doi.org/10.1103/RevModPhys.93.015003}{\emph{Rev. Mod.
  Phys.} {\bfseries 93} (2021) 015003}
  [\href{https://arxiv.org/abs/1908.03430}{{\ttfamily 1908.03430}}].

\bibitem{Brax:2021wcv}
P.~Brax, S.~Casas, H.~Desmond and B.~Elder, \emph{{Testing Screened Modified
  Gravity}}, \href{https://doi.org/10.3390/universe8010011}{\emph{Universe}
  {\bfseries 8} (2021) 11} [\href{https://arxiv.org/abs/2201.10817}{{\ttfamily
  2201.10817}}].

\bibitem{Sakstein:2019fmf}
J.~Sakstein and M.~Trodden, \emph{{Early Dark Energy from Massive Neutrinos as
  a Natural Resolution of the Hubble Tension}},
  \href{https://doi.org/10.1103/PhysRevLett.124.161301}{\emph{Phys. Rev. Lett.}
  {\bfseries 124} (2020) 161301}
  [\href{https://arxiv.org/abs/1911.11760}{{\ttfamily 1911.11760}}].

\bibitem{CarrilloGonzalez:2020oac}
M.~Carrillo~Gonz\'alez, Q.~Liang, J.~Sakstein and M.~Trodden,
  \emph{{Neutrino-Assisted Early Dark Energy: Theory and Cosmology}},
  \href{https://doi.org/10.1088/1475-7516/2021/04/063}{\emph{JCAP} {\bfseries
  04} (2021) 063} [\href{https://arxiv.org/abs/2011.09895}{{\ttfamily
  2011.09895}}].

\bibitem{CarrilloGonzalez:2023lma}
M.~Carrillo~Gonz\'alez, Q.~Liang, J.~Sakstein and M.~Trodden,
  \emph{{Neutrino-Assisted Early Dark Energy is a Natural Resolution of the
  Hubble Tension}},  \href{https://arxiv.org/abs/2302.09091}{{\ttfamily
  2302.09091}}.

\bibitem{Chen:2014oda}
Y.-J.~Chen, W.~Tham, D.~Krause, D.~Lopez, E.~Fischbach and R.~Decca,
  \emph{{Stronger Limits on Hypothetical Yukawa Interactions in the
  30\textendash8000~nm Range}},
  \href{https://doi.org/10.1103/PhysRevLett.116.221102}{\emph{Phys. Rev. Lett.}
  {\bfseries 116} (2016) 221102}
  [\href{https://arxiv.org/abs/1410.7267}{{\ttfamily 1410.7267}}].

\bibitem{Lee:2020zjt}
J.~Lee, E.~Adelberger, T.~Cook, S.~Fleischer and B.~Heckel, \emph{{New Test of
  the Gravitational $1/r^2$ Law at Separations down to 52 $\mu$m}},
  \href{https://doi.org/10.1103/PhysRevLett.124.101101}{\emph{Phys. Rev. Lett.}
  {\bfseries 124} (2020) 101101}
  [\href{https://arxiv.org/abs/2002.11761}{{\ttfamily 2002.11761}}].

\bibitem{Kapner:2006si}
D.J.~Kapner, T.S.~Cook, E.G.~Adelberger, J.H.~Gundlach, B.R.~Heckel, C.D.~Hoyle
  et~al., \emph{{Tests of the gravitational inverse-square law below the
  dark-energy length scale}},
  \href{https://doi.org/10.1103/PhysRevLett.98.021101}{\emph{Phys. Rev. Lett.}
  {\bfseries 98} (2007) 021101}
  [\href{https://arxiv.org/abs/hep-ph/0611184}{{\ttfamily hep-ph/0611184}}].

\bibitem{Tan:2020vpf}
W.-H.~Tan et~al., \emph{{Improvement for Testing the Gravitational
  Inverse-Square Law at the Submillimeter Range}},
  \href{https://doi.org/10.1103/PhysRevLett.124.051301}{\emph{Phys. Rev. Lett.}
  {\bfseries 124} (2020) 051301}.

\bibitem{Hoskins:1985tn}
J.K.~Hoskins, R.D.~Newman, R.~Spero and J.~Schultz, \emph{{Experimental tests
  of the gravitational inverse square law for mass separations from 2-cm to
  105-cm}}, \href{https://doi.org/10.1103/PhysRevD.32.3084}{\emph{Phys. Rev. D}
  {\bfseries 32} (1985) 3084}.

\bibitem{Smith:1999cr}
G.L.~Smith, C.D.~Hoyle, J.H.~Gundlach, E.G.~Adelberger, B.R.~Heckel and
  H.E.~Swanson, \emph{{Short range tests of the equivalence principle}},
  \href{https://doi.org/10.1103/PhysRevD.61.022001}{\emph{Phys. Rev. D}
  {\bfseries 61} (2000) 022001}.

\bibitem{Berge:2017ovy}
J.~Berg\'e, P.~Brax, G.~M\'etris, M.~Pernot-Borr\`as, P.~Touboul and
  J.-P.~Uzan, \emph{{MICROSCOPE Mission: First Constraints on the Violation of
  the Weak Equivalence Principle by a Light Scalar Dilaton}},
  \href{https://doi.org/10.1103/PhysRevLett.120.141101}{\emph{Phys. Rev. Lett.}
  {\bfseries 120} (2018) 141101}
  [\href{https://arxiv.org/abs/1712.00483}{{\ttfamily 1712.00483}}].

\bibitem{Bottaro:2023gep}
S.~Bottaro, A.~Caputo, G.~Raffelt and E.~Vitagliano, \emph{{Stellar limits on
  scalars from electron-nucleus bremsstrahlung}},
  \href{https://doi.org/10.1088/1475-7516/2023/07/071}{\emph{JCAP} {\bfseries
  07} (2023) 071} [\href{https://arxiv.org/abs/2303.00778}{{\ttfamily
  2303.00778}}].

\bibitem{Uzan:2010pm}
J.-P.~Uzan, \emph{{Varying Constants, Gravitation and Cosmology}},
  \href{https://doi.org/10.12942/lrr-2011-2}{\emph{Living Rev. Rel.} {\bfseries
  14} (2011) 2} [\href{https://arxiv.org/abs/1009.5514}{{\ttfamily
  1009.5514}}].

\bibitem{Planck:2018vyg}
{\scshape Planck} collaboration, \emph{{Planck 2018 results. VI. Cosmological
  parameters}},
  \href{https://doi.org/10.1051/0004-6361/201833910}{\emph{Astron. Astrophys.}
  {\bfseries 641} (2020) A6}
  [\href{https://arxiv.org/abs/1807.06209}{{\ttfamily 1807.06209}}].

\bibitem{Will:2014kxa}
C.M.~Will, \emph{{The Confrontation between General Relativity and
  Experiment}}, \href{https://doi.org/10.12942/lrr-2014-4}{\emph{Living Rev.
  Rel.} {\bfseries 17} (2014) 4}
  [\href{https://arxiv.org/abs/1403.7377}{{\ttfamily 1403.7377}}].

\bibitem{Bertotti:2003rm}
B.~Bertotti, L.~Iess and P.~Tortora, \emph{{A test of general relativity using
  radio links with the Cassini spacecraft}},
  \href{https://doi.org/10.1038/nature01997}{\emph{Nature} {\bfseries 425}
  (2003) 374}.

\bibitem{Joyce:2014kja}
A.~Joyce, B.~Jain, J.~Khoury and M.~Trodden, \emph{{Beyond the Cosmological
  Standard Model}},
  \href{https://doi.org/10.1016/j.physrep.2014.12.002}{\emph{Phys. Rept.}
  {\bfseries 568} (2015) 1} [\href{https://arxiv.org/abs/1407.0059}{{\ttfamily
  1407.0059}}].

\bibitem{Sakstein:2020axg}
J.~Sakstein, D.~Croon, S.D.~McDermott, M.C.~Straight and E.J.~Baxter,
  \emph{{Beyond the Standard Model Explanations of GW190521}},
  \href{https://doi.org/10.1103/PhysRevLett.125.261105}{\emph{Phys. Rev. Lett.}
  {\bfseries 125} (2020) 261105}
  [\href{https://arxiv.org/abs/2009.01213}{{\ttfamily 2009.01213}}].

\bibitem{Reyes:2024lzu}
C.~Reyes and J.~Sakstein, \emph{{Parametrized post-Tolman-Oppenheimer-Volkoff
  framework for screened modified gravity with an application to the secondary
  component of GW190814}},
  \href{https://doi.org/10.1103/PhysRevD.109.084080}{\emph{Phys. Rev. D}
  {\bfseries 109} (2024) 084080}
  [\href{https://arxiv.org/abs/2403.03399}{{\ttfamily 2403.03399}}].

\bibitem{Silva:2017uqg}
H.O.~Silva, J.~Sakstein, L.~Gualtieri, T.P.~Sotiriou and E.~Berti,
  \emph{{Spontaneous scalarization of black holes and compact stars from a
  Gauss-Bonnet coupling}},
  \href{https://doi.org/10.1103/PhysRevLett.120.131104}{\emph{Phys. Rev. Lett.}
  {\bfseries 120} (2018) 131104}
  [\href{https://arxiv.org/abs/1711.02080}{{\ttfamily 1711.02080}}].

\bibitem{Doneva:2022ewd}
D.D.~Doneva, F.M.~Ramazano\u{g}lu, H.O.~Silva, T.P.~Sotiriou and
  S.S.~Yazadjiev, \emph{{Spontaneous scalarization}},
  \href{https://doi.org/10.1103/RevModPhys.96.015004}{\emph{Rev. Mod. Phys.}
  {\bfseries 96} (2024) 015004}
  [\href{https://arxiv.org/abs/2211.01766}{{\ttfamily 2211.01766}}].

\bibitem{Akmal:1998cf}
A.~Akmal, V.R.~Pandharipande and D.G.~Ravenhall, \emph{{The Equation of state
  of nucleon matter and neutron star structure}},
  \href{https://doi.org/10.1103/PhysRevC.58.1804}{\emph{Phys. Rev. C}
  {\bfseries 58} (1998) 1804}
  [\href{https://arxiv.org/abs/nucl-th/9804027}{{\ttfamily nucl-th/9804027}}].

\bibitem{Damour:1992we}
T.~Damour and G.~Esposito-Farese, \emph{{Tensor multiscalar theories of
  gravitation}}, \href{https://doi.org/10.1088/0264-9381/9/9/015}{\emph{Class.
  Quant. Grav.} {\bfseries 9} (1992) 2093}.

\bibitem{Esposito-Farese:2004azw}
G.~Esposito-Farese, \emph{{Tests of scalar-tensor gravity}},
  \href{https://doi.org/10.1063/1.1835173}{\emph{AIP Conf. Proc.} {\bfseries
  736} (2004) 35} [\href{https://arxiv.org/abs/gr-qc/0409081}{{\ttfamily
  gr-qc/0409081}}].

\bibitem{Poulin:2023lkg}
V.~Poulin, T.L.~Smith and T.~Karwal, \emph{{The Ups and Downs of Early Dark
  Energy solutions to the Hubble tension: A review of models, hints and
  constraints circa 2023}},
  \href{https://doi.org/10.1016/j.dark.2023.101348}{\emph{Phys. Dark Univ.}
  {\bfseries 42} (2023) 101348}
  [\href{https://arxiv.org/abs/2302.09032}{{\ttfamily 2302.09032}}].

\bibitem{Rudelius:2022gyu}
T.~Rudelius, \emph{{Constraints on early dark energy from the axion weak
  gravity conjecture}},
  \href{https://doi.org/10.1088/1475-7516/2023/01/014}{\emph{JCAP} {\bfseries
  01} (2023) 014} [\href{https://arxiv.org/abs/2203.05575}{{\ttfamily
  2203.05575}}].

\bibitem{Ramadan:2023ivw}
O.F.~Ramadan, T.~Karwal and J.~Sakstein, \emph{{Attractive proposal for
  resolving the Hubble tension: Dynamical attractors that unify early and late
  dark energy}}, \href{https://doi.org/10.1103/PhysRevD.109.063525}{\emph{Phys.
  Rev. D} {\bfseries 109} (2024) 063525}
  [\href{https://arxiv.org/abs/2309.08082}{{\ttfamily 2309.08082}}].

\bibitem{Abbott:1984qf}
L.F.~Abbott, \emph{{A Mechanism for Reducing the Value of the Cosmological
  Constant}}, \href{https://doi.org/10.1016/0370-2693(85)90459-9}{\emph{Phys.
  Lett. B} {\bfseries 150} (1985) 427}.

\bibitem{Graham:2015cka}
P.W.~Graham, D.E.~Kaplan and S.~Rajendran, \emph{{Cosmological Relaxation of
  the Electroweak Scale}},
  \href{https://doi.org/10.1103/PhysRevLett.115.221801}{\emph{Phys. Rev. Lett.}
  {\bfseries 115} (2015) 221801}
  [\href{https://arxiv.org/abs/1504.07551}{{\ttfamily 1504.07551}}].

\bibitem{Hindmarsh1983ODEPACK}
A.C.~Hindmarsh, \emph{{ODEPACK, A Systematized Collection of ODE Solvers}},
  {\emph{IMACS Transactions on Scientific Computation} {\bfseries 1} (1983)
  55}.

\bibitem{2019ascl.soft05021H}
A.C.~{Hindmarsh}, ``{ODEPACK: Ordinary differential equation solver library}.''
  Astrophysics Source Code Library, record ascl:1905.021, May, 2019.

\bibitem{Kierzenka2001ABS}
J.~Kierzenka and L.F.~Shampine, \emph{A bvp solver based on residual control
  and the maltab pse}, \href{https://doi.org/10.1145/502800.502801}{\emph{ACM
  Trans. Math. Softw.} {\bfseries 27} (2001) 299–316}.

\bibitem{Ascher1995}
U.M.~Ascher, R.~Matheij and R.D.~Russell, \emph{Numerical Solution of Boundary
  Value Problems for Ordinary Differential Equations}, Society for Industrial
  and Applied Mathematics, Philadelphia (1995),
  \href{https://doi.org/10.1137/1.9781611971231}{10.1137/1.9781611971231}.

\bibitem{Gao:2012guu}
F.~Gao and L.~Han, \emph{{Implementing the Nelder-Mead simplex algorithm
  with~adaptive parameters}},
  \href{https://doi.org/10.1007/s10589-010-9329-3}{\emph{Comput. Optim. Appl.}
  {\bfseries 51} (2012) 259}.

\end{thebibliography}\endgroup
\end{document}